# GROWTH, INDUSTRIAL EXTERNALITY, PROSPECT DYNAMICS, AND WELL-BEING ON MARKETS

Emmanuel CHAUVET

Laboratoire de Physique Théorique Fondamentale en PACA
Département de Sciences Economiques
chauvet-emmanuel@bbox.fr

**Abstract** Functions or "functionings" enable to give a structure to any activity and their combinations constitute the capabilities which characterize economic assets such as work utility. The basic law of supply and demand naturally emerges from that structure while integrating this utility within frames of reference in which conditions of growth and associated inflation are identified in the exchange mechanisms. Growth sustainability is built step by step taking into account functional and organizational requirements which are followed through a project up to a product delivery with different levels of externalities. Entering the market through that structure leads to designing basic equations of its dynamics and to finding canonical solutions, or particular equilibria, after specifying the notion of maturity introduced in order to refine the basic model. This approach allows to tackle behavioral foundations of Prospect Theory through a generalization of its probability weighting function for rationality analyses which apply to Western, Educated, Industrialized, Rich, and Democratic societies as well as to the poorest ones. The nature of reality and well-being appears then as closely related to the relative satisfaction reached on the market, as it can be conceived by an agent, according to business cycles; this reality being the result of the complementary systems that govern human mind as structured by rational psychologists. The final concepts of growth integrate and extend the maturity part of the behavioral model into virtuous or erroneous sustainability.

1. Introduction

"The concept of 'functionings', which has distinctly Aristotelian roots, reflects the various things a person may value doing or being" (Sen 2001). This is the Sen's side taken for the development of a functional economics initially based on engineering tools (Chauvet 2013). On the labor market both know-how and inter-personal skills may be valued by employers as well as by employees of any firm. The structuring of economic activities into functions enables to encompass "being" and "doing" in the same model: working is a means of gathering both notions that provide any agent with the position he or she deserves among the other workers of a community. Beyond decision-making which is instantaneous, although it has to be prepared and it can lead to long-lasting activities, action always comes within a time evolution: for instance, in physics it is the product of energy or work by the duration of the measured process. So, time must be introduced as the money standard which applies to economic efficiency measurement whatever the currencies that are used by the firms on the markets.

Time resource is scarce especially when an agent grows old what often goes with higher responsibilities and thus a transition from action to decision-making. According to Sen a "person's capability refers to the alternative combinations of functionings that are feasible for her to achieve" and to a certain extent such capability increases also with experience while an agent gets over the steps of that transition. Eventually this analysis can be applied within a formalism showing the intricacy of capability and time for the definition of work utility.

This work utility can be represented as a curve but it has no known origin or rather its beginning is lost in the course of evolution. So as to use it the economist has to "climb on the bandwagon" and fix a point that will be the origin of his study. Hidden in this representation lies the notion of gauge, in terms of which economic phenomena can be analyzed. Finding a reference point introduces behavioral principles and in particular the one of the framing effects of the Prospect Theory (Kahneman and Tversky 1979). Within a given frame, work utilities can be exploited to go deeper into the basic law of supply and demand which assumes a vast number of equilibria under its functional form (Chauvet 2013). A fine study can show the high degree of interdependence of the agents that are involved in a transaction or an exchange, those agents often being contractually bound. The "buy and forget" behavior is no longer appropriate when a complex purchase is made which requires a learning period and the resort to a support from the supplier or a skilled subcontractor. Moreover, this period and the efficiency of the support prove to be necessary conditions for a global growth within the finest framework of the law of supply and demand. The relation between this microeconomic growth principle and inflation can then be emphasized.

Going deeper into the functional structure of a project prospects and refining the payoff concept – as the ratio of the adaptive utility to the capability of agents – allow a rational point of view resorting neither to currencies nor to a complex Game Theory. But this payoff relying on adaptive utility applies essentially to marketing whereas value is built along with the creation of effective functions within the project intended to design, develop and provide a given contractual customer with a solution, or to launch a product, good or service on a specific market. These functions are associated with energies and physical work.

So functionings and functions show a notional continuity from work or energy provided by workers or by means of production and eventually developed by products, to a strategic adaptive utility which is used by market makers as a means of gathering in the same functional concept all the cognitive charge in relation with the various aspects of the firm management and of its goods or services marketing. Although work and energy, on the one hand, and adaptive utility, on the other hand, may be distinguished analytically, both notions can find an economic interest in being unified and identified one to the other in some circumstances. Such interest appears clearly for projects which are highly visible from the hierarchy of the firm. Anticipating those projects success, the associated value is seen as the aggregation of the ones of all developed function. The use of the capacity formalism (Choquet 1954) according to the Prospect Theory (Tversky and Kahneman 1992) allows to refine the expression of this value within the valorization of outcomes – energy or work – under organizational and technical constraints of profitability. These constraints can be shown as the origin of the relation between inflation and unemployment as noticed Phillips (1958) through an empirical study.

Whether it comes from a bid process in order to win a contract for the achievement of a project for a customer or it consists of an investment of the firm to produce a good or service for the mass consumption, a budget must be drawn up which relies on the human and skills capital of the company and on the information required to manage the project. The complexity of industrial arrangements can be described through matrix formalism, from the tenderer's submission and anticipated design and architecture of the solution proposed to the customer, to its delivery after completion in spite of possible unexpected externalities.

Once established the industrial processes and organization for providing the customers with solutions to contractual requirements or with goods and services, the market dynamics can be studied for a supply policy, respectively for the demand, through two fundamental equations taking account of the framing approach. Particular situations can be described: pure speculation when no value is exchanged on the market but which characterizes innovative ideas including negative externalities processing; equilibrium when the payoff function is constant; and the general case according to which the solutions of the market dynamics equations show similarities while choosing constraints on their parameters. This choice based on maturity and complexity or capacity restrictive relations leads to define general conditions for exchanges to be possible.

These conditions introduced in the behavioral model of Tversky and Kahneman (1992) enable to widen their fourfold pattern of decision under risk – initially applicable to money managers and graduate students from great American universities – to the perception of markets the consequence of which is a specific behavior according to business cycles. The initial application of the fourfold pattern has been mainly exploited within Western, Educated, Industrialized, Rich, and Democratic (WEIRD) societies and it has been shown (Henrich et al. 2010) that generalizing about humans' behavior from WEIRD behavioral standards is a real bias that deserves a particular attention. So, in search of universality, the initial pattern can be extended to poor economics (Banerjee and Duflo 2012) in using the probability weighting function of Tversky and Kahneman as an operator for behavioral analysis and for the determination of the well-being conditions according to the concept of market. Reality then appears as a mental equilibrium induced by the functioning of the two systems which allow the description of a part of the human behavior (Kahneman 2011).

Built upon this dualist model of welfare, new conditions characterize two ways for the capital to grow: the virtuous growth in positive maturity economic sectors and the erroneous growth associated with unmatured markets.

2. Standard Law of Supply and Demand

*2.1. Work Duration as a Money Standard*

Let $\rho$ be a work duration that can find a meaning in terms of money through the notions of wage or hourly rate. Then $\rho$ can be considered as the analog of a cardinal measure of a work utility when it can be admitted that the corresponding activity has a real economic value through functionings or functions creation. Such a measure can also be understood as a way for an agent to be active on markets: on the labor market when he or she produces goods and services under the brand name of a firm or as a self-employed worker; on the markets of all kinds of products where he or she has to buy so as to satisfy his or her needs for food, transportation, lodging, entertainment and so on. In order to avoid any misuse of existing meanings relying on a vast amount of works on the principle of cardinal utility, $\rho$ is named "time cost" while a salaried employment is a means of paying and enjoying all what life can offer.

As a builder of goods and services, the agent provides functions that are directly or indirectly used for the production of increasing complexity products, those functions relying also on raw materials that must be converted and on machines and tools which enable the creation of value thanks to high skilled workers. Those functions are "priced" as shown in figure 1. The "price" of the function $f_m$ which is found between $\rho_i$ and $\rho_j$ depends on the practices of its provider.

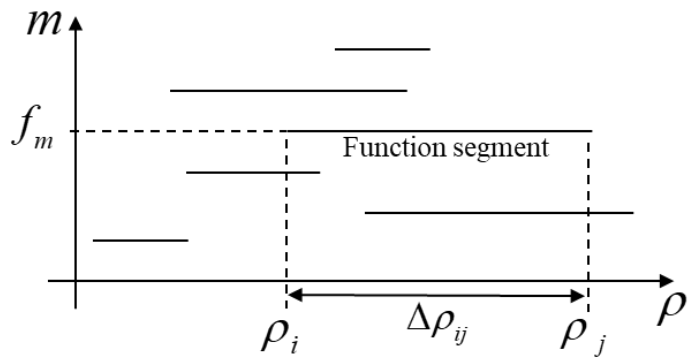

**Fig. 1.** Function pricing

## 2.2. Functional Capability and Work Utility

But the skills which are developed by the agent are not even among a population of workers and it is necessary to introduce a means of making a distinction between their capabilities, as the ordinal utility enables to distinguish relative perceived values by an order of preference. Thus, a capability can be established as the ratio of the variation of $m$, the number of provided functions plunged into the set of real numbers, to the variation of the work duration: the greater the ratio, the higher the functional capability $C$. The global work utility is then:

$$W_u = \rho C = \rho \frac{dm}{d\rho} \qquad (1)$$

Hence there are two ways to compare work powers as there are two ways to structure utility: the cardinal one $\rho$ that gives a measurable scale and the ordinal one which is the derivative of $m$ according to $\rho$ and enables also to take into account preferences. Those preferences are based on decisions about which functional practices are selected through the work utility estimate.

## 2.3. Supply and Demand Within the Framing Effects

The work product, in functional terms, corresponds to the function segments crossed by the work utility curve which is linear in a first approximation due to a framing effect (figure 2). The number of functions selected in that way – according to preference of practices – is the complexity (c) of the work product. As it can be considered as a supply (s) or as a demand (d), it seems relevant to distinguish two complexities, one for each economic prospect that is conceivable on a market. The work utilities are such that:

$$W_{us} = \rho_s \left(\frac{dm}{d\rho}\right)_s = c_s \quad ; \quad W_{ud} = \rho_d \left[-\left(\frac{dm}{d\rho}\right)_d\right] = c_d \qquad (2)$$

The figure 2 shows how to understand the two kinds of work utilities and their connection with the functional complexity of the work product that is supposed to be exchanged in the supply and demand frame but it does not provide the way how the exchange price is fixed.

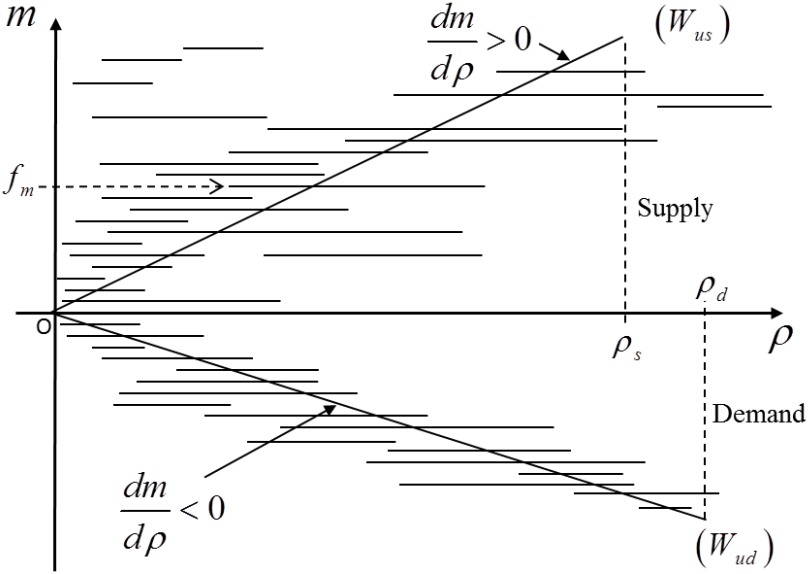

**Fig. 2.** Work utility within a specific frame

In this figure 2, both behaviors of supply and of demand are presented simultaneously but it must be clear that each linear approximation is performed by a different agent although exchanges may take place between entities belonging to the same organization or firm.

In a more complex model, the work utility is not linear and can globally involve phases when the agent is a provider of function and others when he or she is a consumer. The same kind of behavior must be understandable also for a firm which has suppliers and customers.

The figure 3 sets out different frames of study that illustrate how the framing effect can be understood. The global one $(O, \rho, m)$ shows the full work utility curve while the frame centered on O' presents a section of the curve where the linear approximation applies in a phase of demand and the frame centered on O'' shows another part where the same kind of approximation also applies in a phase of supply.

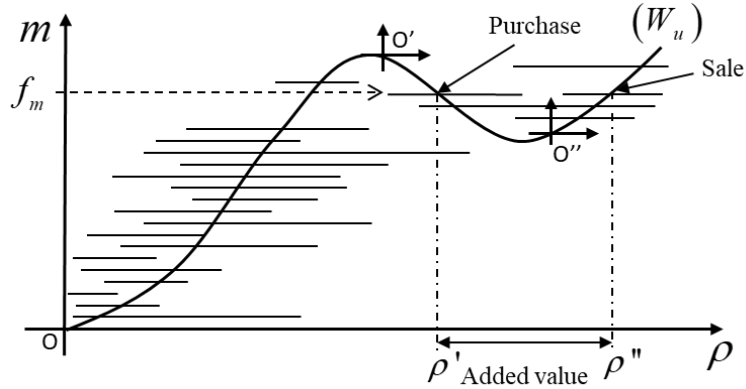

**Fig. 3.** Global work utility curve of an agent

In the global framework, the function $f_m$ is firstly bought at a "price" $\rho'$, is then enhanced, integrated in a product or simply resold under the brand name of the buyer at a price $\rho''$ after value has been added concretely or by the simple fact that a name may have a reputation which deserves a margin that is worth a higher level of warranty.

*2.4. Supply and Demand and Capital as a Framework*

Applying a financial reasoning leads to link capital and work utility while admitting that all elementary functions are realized – designed and prepared in an intellectual way and then industrialized – or performed more concretely at the same "time cost". The capital K can be written:

$$K = \rho m \tag{3}$$

So, after differentiation in order to account for the great variety of functions and "prices":

$$dK = \rho dm + m d\rho \tag{4}$$

and according to the expressions of the work utilities:

$$\rho \frac{dm}{d\rho} = \frac{dK}{d\rho} - m = (\pm)c \qquad (5)$$

After integration – of the work utility – for the supply with Os as the reference point:

$$\rho_s = \rho_{Os} \exp\left(\frac{m_s}{c_s} - \frac{m_{Os}}{c_{Os}}\right) \qquad (6)$$

and for the demand:

$$\rho_d = \rho_{Od} \exp\left(\frac{m_{Od}}{c_{Od}} - \frac{m_d}{c_d}\right) \qquad (7)$$

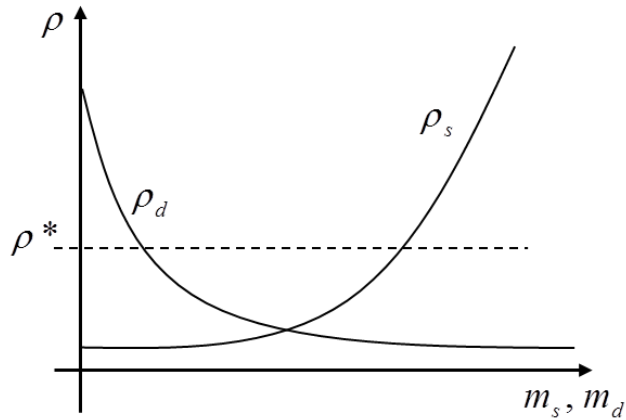

**Fig. 4.** Supply and demand curves

These last results have already been established by Chauvet (2013) for an explanation of the law of functional supply and demand as illustrated by the figure 4.

From the viewpoint of the capital, it can be added for the supply that:

$$\frac{dK_s}{d\rho_s} = c_s \left[\ln\left(\frac{\rho_s}{\rho_{Os}}\right) + \frac{m_{Os}}{c_{Os}} + 1\right] \qquad (8)$$

$$K_s = c_s \rho_s \left[ \ln \frac{\rho_s}{\rho_{Os}} + \frac{m_{Os}}{c_{Os}} \right] + K_{Os} \qquad (9)$$

And for the demand, the capital follows the law:

$$K_d = c_d \rho_d \left[ \frac{m_{Od}}{c_{Od}} - \ln \frac{\rho_d}{\rho_{Od}} \right] + K_{Od} \qquad (10)$$

Thus, when the exchange takes place and when there is an agreement on the price $\rho^*$, the global capital is:

$$K = (K_d + K_s)_{\rho_d, \rho_s = \rho^*} \qquad (11)$$

$$K = \rho^* \left[ \ln \left( \left( \frac{\rho^*}{\rho_{Os}} \right)^{c_s} \left( \frac{\rho_{Od}}{\rho^*} \right)^{c_d} \right) + \frac{c_s}{c_{Os}} m_{Os} + \frac{c_d}{c_{Od}} m_{Od} \right] + K_{O(s,d)} \qquad (12)$$

*2.5. Conditions of growth*

In the simple case where the complexity imagined by the buyer is equal to the complexity designed and developed by the provider, there comes:

$$\Delta K = K - K_{O(s,d)} = c\rho^* \left[ \ln \left( \frac{\rho_{Od}}{\rho_{Os}} \right) + \frac{m_{Os}}{c_{Os}} + \frac{m_{Od}}{c_{Od}} \right] \qquad (13)$$

The global capital of the buyer and of the seller increases whenever this condition is verified:

$$r = \frac{\rho_{Od}}{\rho_{Os}} > \exp \left( -\frac{m_{Os}}{c_{Os}} - \frac{m_{Od}}{c_{Od}} \right) \qquad (14)$$

This capital growth depends naturally on the complexity and on the cost – or work time cost – and then on the price of the product that is exchanged but also essentially on the frame, the references of which are the points Os and Od.

In the light of the equilibrium $(\rho_s, c_s) = (\rho_d, c_d) = (\rho^*, c)$ the ratio r is the opportunity for bargaining the delivery conditions of the product that will be exchanged between the supplier and the buyer: those conditions consist essentially in the level of support that will be provided for enabling the takeover of the product after the purchase. The support duration is generally included in the offer but the buyer often forgets that he or she will also spend more or less time on appealing to it. This bargaining depends on the past experiences of both the supplier and the customer but may not be fully appropriate to the new situation: although the takeover has been prepared, the transition which characterizes the exchange is always surrounded with a considerable degree of uncertainty.

When both supply and demand of the references are saturated, $(c_{Os}, c_{Od}) = (m_{Os}, m_{Od})$, the capital growth is positive if the time cost invested by the demand is more than r = 13.5% of the one paid to the supply.

$$r = \frac{\rho_{Od}}{\rho_{Os}} = \frac{\rho_u t_{Od}}{\rho_u t_{Os}} = \frac{t_{Od}}{t_{Os}} > 13.5\% \qquad (15)$$

In other terms for the growth to be positive, under the hypothesis of a uniform cost $\rho_u$, when an agent buys a good or a service he or she must accept to spend on the takeover of his purchase more than 13.5% of the time used by the seller for the production and the distribution of the good or service.

Out of saturation, $c_{Os} < m_{Os}$, the supplier selected a restricted number of functions for the production and distribution; or $c_{Od} < m_{Od}$, the buyer focused his or her purchase only on the functions he or she needs; the ratio of time costs is reduced for several reasons: the supplier specialized and the buyer paid only for what he was in need of. As a consequence, the condition of growth corresponds to a lower ratio r and the required time spent on the takeover is relatively shorter thanks to a better fit.

Both conditions $c_{Os} < m_{Os}$ and $c_{Od} < m_{Od}$ mean that the reference points have been reached while the numbers of functions supplied $m_{Os}$ and bought $m_{Od}$ were above the respective complexities to which the supplier and his or her customer had paid a particular attention. It follows that, r being constant, the growth is all the more important since the gap – between the number of functions of a product and its perceived complexity – is large: generally when someone buys a complex system, a car for instance, he or she often does not mind all the vast number of technical functions which contribute to the few basic functions that are eventually required for a relatively easy use. So, an efficient complexity management is also a means of generating growth.

However, this growth relies on information: a better knowledge of the markets or of the functions available for production or for consumption but also Big Data analysis or analytics and processing tools in order to create Artificial Intelligence for complexity management.

Within industrial exchanges of complex systems or machine tools delivered to firms or factories for their production lines, the cases $c_{Os} > m_{Os}$ and $c_{Od} > m_{Od}$ can appear when the numbers of functions supplied $m_{Os}$ and bought $m_{Od}$ hide the real complexities that must be taken in charge. Actually, these functions show decompositions into sub-functions the control of which is decisive for a system takeover and operation and requires high-skilled and trained workers. In these cases, the ratio r of the required time costs – for the growth to be possible – increases with the complexities showing how much a firm can be dependent upon the systems, the corresponding training and support it has been provided with and upon its providers, subcontractors or industrial partners: the time [cost] that must be spent by the purchaser for the takeover is relatively high in comparison with the time spent by the provider for the production of the purchase. The explanation is that the customer has gone through a long process of specification of requirements prior to this production and that he or she needs training and support for the use of the resulting complex system.

Considering the labor market and a kind of standardization of the time costs for given employments and skills, whatever the business involved in the race for a global capital growth, this one is all the greater since r is large in comparison with the exponential term. This expresses the fact that the higher the relative level of skill of the demand, the less training is needed for the purchase takeover. Furthermore, the capital which increases all the more so since $c\rho$ is great shows that growth is accelerated by an increase in complexity or in time cost or in both.

*2.6. Global capital growth and inflation*

For the global capital to grow within an exchange some conditions on the reference frames must be realized. Among others, a deeper study of the constraint on the ratio r enables to understand the origin of the inflation – or of the deflation – related to an "ex ante" increase of complexity: when $m_{Os}$ and $m_{Od}$ are kept constant while $c_{Os}$ and/or $c_{Od}$ rise, r also increases what induces that $\rho_{Od}$ gets greater or $\rho_{Os}$ diminishes or both, the most probable being that $\rho_{Os}$ stays constant or grows less than $\rho_{Od}$. Thus logically an increasing complexity should lead at first to a marked increase in the time cost of the demand and, through the adjustment of salaries in a given economic sector, to a global inflation in this sector: when the perceived complexities decrease, because of and after a rise in competence on the corresponding range of products, the constraint on r is relaxed what implies that $\rho_{Os}$ can in turn increase more than $\rho_{Od}$. This reasoning relies essentially on the evolution and learning of practices, the complexity of which depends on its perception [on the market] and so on cognitive criteria: once a practice is well understood and widespread it can become a standard of work and behavior. Thus, change driven by innovation to temporarily higher perceived complexity is necessary for a sustainable economic growth.

If such a standard is not challenged by changes to a higher complexity, it can lead to deflation: the demand would have no reason to replace its practices or its assets with equivalent ones of flat perceived complexity characterizing a low innovation context, so $\rho_{Od}$ would decrease until the ratio r reaches its constraint level and below, $\rho_{Os}$ must be reduced in order to maintain a reasonable growth. Then a solution to avoid deflation at a microeconomic level is planned obsolescence which sustains the demand for replacement of goods and services within this low innovation context.

When $m_{Os}$ and/or $m_{Od}$ increase while a higher level of complexity is developed through innovation, the ratio r can also decrease according to a lower constraint leading to time costs reductions in some circumstances of disinflation when the innovation enables productivity gains. But this decrease in r can also lead to greater time costs when this innovation goes with its widespread use thanks to a great deal of consumers or suppliers willing to adopt new standards and behaviors.

3. Payoff and value

Let $F$ be a set of all the conceivable functions in a given field of economic activities:

$$F = \{f_i\}_{1 \leq i \leq m_F} \tag{16}$$

with $m_F$ the cardinality of $F$. This set constitutes a strategy for some marketing purpose. As designed, each of its functions is supposed to develop the energy or work or adaptive utility $E_i$ (Chauvet 2013). More precisely, $F$ is also written:

$$F = F_s \cup F_d = \{f_i^s\}_{1 \leq i \leq \lceil m_s \rceil} \cup \{f_i^d\}_{1 \leq i \leq \lceil |m_d| \rceil} \tag{17}$$

The set $F_s$ represents the functions that the firm is ready to supply and $F_d$, the functions that the customers are waiting for, such that $F_s \cap F_d \neq \emptyset$. $\lceil m_s \rceil$ and $\lceil |m_d| \rceil$, are the ceiling functions respectively of $m_s$ and $|m_d|$.

On the one hand, $F$ can be reached through the market analysis and studied theoretically in universities or schools and also by professionals who have to maintain a high level of knowledge required to manage the evolution of their practices in relation with more technical teams of skilled workers and colleagues. They also need $F$ for reporting to their superiors who appreciate understanding simply the way how their business grows, when it grows on a market segment, or the reason why a strategy does not fit the requirements of their prospective customers.

On the other hand, $F$ has to be regularly "tuned" according to the physical, technical and economic constraints that must be taken into account for the development of products or services according to another set $F_R$:

$$F_R = \{g_i\}_{1 \leq i \leq m_R} \tag{18}$$

This set is made up of all the functions that are really performed by the solutions brought to the market by a firm through a project $S$. Each function of this set yields the outcome $E_i^R$ which can be considered as a work (or energy).

*3.1. The Payoff*

The payoff function can then be defined as:

$$P = \frac{1}{C} \sum_{i=1}^{i=m} E_i \qquad (19)$$

where $C = C_s, C_d$, $m = \lceil m_s \rceil, \lceil |m_d| \rceil$ and $E_i = E_i^s, E_i^d$ whether $P$ is written for the supply or for the demand. This definition expresses the fact that the payoff to a firm is proportional to the adaptive utility or effective work developed by its employees but in inverse proportion to their capability needed to develop such a utility. As asserted by Chauvet (2013) the capability is also an indirect measure of the uncertainty, hence its converse is a characterization of the certainty and the payoff is all the more important since the certainty of the outcomes $E_i$'s is great and gives strength to a strategy. In other caricatured words the firm functioning requires employees "who do the job" with the just necessary capabilities and the ability to find certainties, at least temporarily for the success of a project.

*3.2. Project, Outcomes and Organization*

After Tversky and Kahneman (1992), let $S$ be a finite set of states of a project led by a firm, the subsets of which are known as events and let $E$ be a set of outcomes that can be positive or negative – gains or losses – and not only monetary outcomes. An "uncertain" prospect $g$ is a function from $S$ into $E$ such that:

$$\forall s \in S \quad g(s) = \varepsilon, \quad \varepsilon \in E \qquad (20)$$

The outcomes of $g$ are arranged in increasing order in such a way that for $(A_j)$ a partition of $S$, the prospect is a sequence of pairs $(\varepsilon_j, A_j)$ such that $g$ yields $\varepsilon_j$ when the event $A_j$ occurs.

Applying the formalism of Tversky and Kahneman, the value associated with $g$ is:

$$V(g) = \sum_{j=-m}^{j=n} \pi_j v(\varepsilon_j) \tag{21}$$

where $v$ is the way how an outcome is valued and $\pi_j$ is computed according to the concept of capacity $W$ (Choquet 1954) which generalizes the standard notion of probability on a set of events: $W(\varnothing) = 0$, $W(S) = 1$ and $W(A_k) \leq W(A_j)$ whenever $A_k \subset A_j \subset S$.

Here is considered a real event $A$ concerning all the agents committed to producing the set of functions $F_R$ in the framework of the project: each $g_i$, as a service, is performed through the collaboration of agents of the firm, according to events, or as a good, it provides a functionality used by the consumer, customer of the firm. For the whole project, as it is a collective work, $A = S$ and the associated value is:

$$V = \sum_{i=1}^{m_R} V(g_i) = \sum_{i=1}^{m_R} \sum_{j=-m}^{j=n} \pi_{i,j} v(\varepsilon_{i,j}) \tag{22}$$

where $n = m = 0$ and $\pi_{i,0} = W(S)$ at the "granularity" of the project. Then:

$$V = \sum_{i=1}^{m_R} W(S) v(E_i^R) = \sum_{i=1}^{m_R} v(E_i^R) \tag{23}$$

*3.3. Project Within the Organization of the Firm and Value*

Nevertheless, contributions inside or outside the firm can be distinguished. The project $S$ is a collaborative work of departments $D_\alpha$ of the firm and each department shares a part of the events that drive it. When it is driven by a market or an innovation strategy, the design of a product relies on the functions $f_i$'s. Thus for this given project it is possible to consider $(D_\alpha)$ as a partition of $S$ while each function or prospect $f_i$ is associated with the subsets or events $(A_{i,j})$ which form another partition of $S$. The outcome of $f_i$ when the event $A_{i,j}$ occurs is $\chi_{i,j}$, a technical component that is necessary to produce the good or service made up with the functions $g_i$'s, target of the project, according to the state of the art on the market. For the considered firm, processes and practices are such that $A_{i,j}$ is adapted to its organization:

$$A_{i,j}^\alpha = A_{i,j} \cap D_\alpha \tag{24}$$

The firm may have to use a subcontractor to get the component $\chi_{i,j}$ or to buy it on the market in order to build its good or service but this specific component is adapted and integrated by the departments of the firm. For the whole $f_i$ it is possible to define $A_i^\alpha$ the contribution of a given department to the production of all the components needed for this function:

$$A_i^\alpha = \bigcup_j A_{i,j}^\alpha = \bigcup_j \left( D_\alpha \cap A_{i,j} \right) \tag{25}$$

The global contribution of the departments to $f_i$ is then:

$$\bigcup_\alpha A_i^\alpha \tag{26}$$

The whole contribution of the firm's organization in terms of events through the project $S$ is then:

$$\bigcup_{i=1}^{m_F} \left( \bigcup_\alpha A_i^\alpha \right) \tag{27}$$

In the case of additivity of the capacities relatively to the functional structure of the product yielded by the project:

$$W\left(\bigcup_{i=1}^{m_F}\left(\bigcup_\alpha A_i^\alpha\right)\right) = \sum_{i=1}^{m_F} W\left(\bigcup_\alpha A_i^\alpha\right) \qquad (28)$$

And when the contributions of the departments $D_\alpha$ to the project $S$ are also additive:

$$\sum_{i=1}^{m_F} W\left(\bigcup_\alpha A_i^\alpha\right) = \sum_{i=1}^{m_F} \sum_\alpha W\left(A_i^\alpha\right) \qquad (29)$$

Founded on a popular belief, the initiation of the project is attributed to a visionary firm leader, whether someone who has seminal ideas or who knows how to federate teams of creative people. In both cases this initiation relies on the combinatorics of all what is possible for a start-up which will "scale fast" or for a major company already having a great organization and a substantial portfolio of profitable projects. The complexity managed, virtually through an intuition, a reliable business plan and a strong financial support for a start-up, or concretely for a powerful and well-established firm, can be assessed as being the organizational cumulative capacity:

$$c_{org} = \sum_{i=1}^{m_F} \sum_\alpha \sum_j W\left(A_{i,j}^\alpha\right) \qquad (30)$$

The nature of the contributions and of the functional or technical structures relatively to the capacities $W$, sub-additive, additive or super-additive, depends on the organization, its synergies and on the practices of the firm. This nature can be analyzed in order to improve margins between the costs or expenses of the project and its profit when its products are delivered to the customers of the firm.

After a combinatorial reduction of the contributions to the technical components of the prospective solution, it appears that these components can be mutualized and reduced to the set $X = \{\chi_k\}$ with $c_{tech} = card(X)$, the number of the components of $X$. A constraint of profitability can then be written:

$$c = \left\lceil \frac{c_{org}}{c_{tech}} \right\rceil \tag{31}$$

Finally, the notion of function is polymorphous and emerges from the diverse concepts used by the multiplicity of contributors of a project to describe a technical system. The work utility $W_u = \rho C = c$, as stated previously, gives pieces of information not only on the complexity involved in the supply or in the demand but also about $c$ understood as the ratio of the cumulative capacity of all the contributors to the number of mutualized technical components, necessary for the production of goods or services provided by the firm according to the project characterized by $S$. The capacity then appears as the product of a capability by a time cost and by an indicator measuring a technical core. It can also be suggested that the global capacity is a measure of the monetary – proportional to $\rho$ – motivation of the teams mobilized in the processes of the firm within a project. The "pernicious" effect of this conclusion is that fixing a constraint of profitability implies also selecting rigorously the project team members in such a way that the complexity will be managed without needless – although motivated – participants. Following the constraint applied to the ratio r – for a global growth to be possible – and according to the new meaning of c, it appears that for a given technical core, an uncontrolled increase in $c_{org}$ can lead to inflation. So fixing a constraint of profitability enables to master the right level of inflation compatible with a sustainable growth, in spite of the consequences on the unemployment: when $c_{org}$ is low, less workers are needed to perform less work events of the project.

At the beginning of the project, the organizational structure is defined for its realization and a costing is worked out while the project team is selected and involved or hired. Generally, the more the amount of contributions is important, the less the outcomes of the functions, that are really produced, are valued or the less the profit. On the other hand, the set of technical components X cannot be reduced or if it is, that would lead to a loss of value, the final product being incomplete compared with what it was expected to be. As a consequence, the value of the outcome $E_i^R$ is expressed by:

$$v\left(E_i^R\right) = \frac{1}{c} E_i^R \rho \tag{32}$$

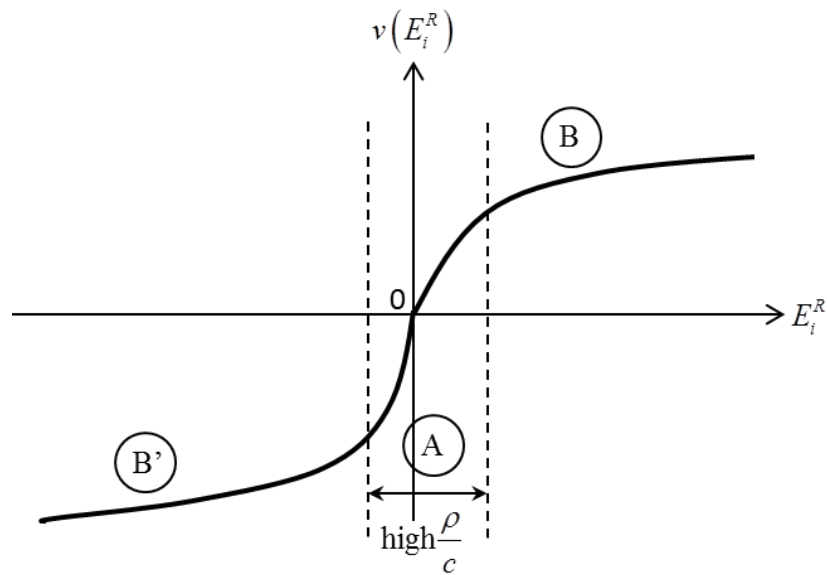

**Fig. 5.** Diminishing returns

The deviation from linearity is due to the ratio of the time cost to the cumulative capacity and may obey the law of diminishing returns to the contributors (part B of the figure 5), like the value of the Prospect Theory (Kahneman & Tversky 1979) verifies the principle of diminishing sensitivity.

When $E_i^R < 0$, the corresponding function constitutes a negative externality and the slope of the value is stiffer (part A of the figure 5) because of the need of a higher time cost to remunerate a small number of experts – such that $c_{org}$ is also low – capable of reducing this negative externality. Within the investigations of those experts, the methods of reduction being specified, the greater the absolute value of $E_i^R$, the more skilled contributors must be involved to mitigate the negative externality: $c_{org}$ increases whereas the time cost is still limited (part B' of the figure 5).

The value produced by the firm, to which is given a concrete expression by a product, is then written:

$$V = \frac{1}{c} \sum_{i=1}^{i=m_R} E_i^R \rho \qquad (33)$$

This expression also accounts for the search for simplicity – less people (low $c_{org}$) to manage a greater number of technical components (high $c_{tech}$) – which is more valued than high a use complexity of a product and value is naturally in direct proportion to the work and its duration necessary to produce a given good or service. Another alternative reasoning based on the Prospect Theory gives an explanation on how the value $v$ can be assessed (Chauvet 2015) through an operational optimization which is fostered as a best practice – for profitability – in the firm.

According to the maturity of the marketing or innovation strategy, $F$ and $F_R$ are close one to the other – especially in an agile project – the latter being built from the technical components yielded upon the design and developments performed through the former enforcement.

Whether it is for the firm a matter of time to market or of compliance with the schedule of a contract, the value produced by a planned project decreases when the availability of the technical components is not obtained within the schedule. This aspect of a project requires distinguishing the contractual framework from the technical facts: the technical course can lead to a successful product although contractual milestones are not gone through as planned in agreement with the customer or as anticipated for a timely introduction on a market. So, a project may be a technical success while it is financially in deficit: the mechanisms of penalty within a contract and of competition on the market imply that the final value of the product is the one of the technical solution reduced drastically by penalties or by the presence of competitors on the target market.

4. Industrial processes under estimates

The industrial processes enable to turn $F$ into $F_R$, the former characterizing mainly the customer's expectation and the latter is the functional basis for the validation of the customer's satisfaction. This validation is formal and relies on tests performed by the provider of the set $F_R$ in order to make sure the corresponding product can be delivered to the market or to a contractual customer. In this last case the final acceptance test and report is generally required to ensure that the contract has been respected. The delivery must occur according to the contractual schedule. Should the good or service, in part or entirely, be delivered beyond the scheduled date of delivery, some penalties can be applied to its provider.

In this paragraph the estimates practiced by the industrial providers show that the initial evaluation is high enough to take into account the main unknown factors conceivable by any contract expert and corporate lawyer. So, it is assumed that any risk involving penalties is covered within the estimates or that the industrial providers have underestimated the expectations of the customer and must be ready to be in deficit for the corresponding contract.

Thus, in what follows, a great effort is made to technically ensure the delivery of what is contractually expected even if it means reducing any uncertainty while selecting the operating modes, the most efficient ones, of the chosen technical components of the delivered solution.

The logic of the following sub-sections is turned towards the customer's satisfaction and, with the appropriate means, towards a profitable activity within a budget shared between firms, then inside a firm between entities, each level of budget allocation being an opportunity of creating value while reducing costs. Margins appear naturally with the uncertainty management.

### 4.1. From Architecture and Design to an Industrial Arrangement

At the bid or anticipated design stage of the project the set $F$ is figured out in a dynamical way according to scenarios of use under budget constraints. An implicit simulation can be formalized by the following doubly stochastic matrix $M_F$ which characterizes the probability $\omega_{i,j}$ for the function $f_i$ to activate [or interact with] the function $f_j$:

$$M_F = \left[\omega_{i,j}\right]_{1 \leq i,j \leq m_F}, \sum_{i=1}^{m_F} \omega_{i,j} = 1, \sum_{j=1}^{m_F} \omega_{i,j} = 1 \tag{34}$$

At this stage based on previous projects or on functional concepts $\omega_{i,j}$ are still vague in the minds of the stakeholders but according to the theorem of Birkhoff (1946) – von Neumann (1953), $M_F$ can be decomposed into permutation matrices $\Pi_\beta$ such that:

$$M_F = \sum_\beta W_\beta \Pi_\beta, \quad \sum_\beta W_\beta = 1, \quad W_\beta \geq 0, \quad \Pi_\beta = \left[\omega_{i,j}^\beta\right]_{1 \leq i, j \leq m_F} \tag{35}$$

This decomposition depends on the [know-how of the] firms and on their target customers and so is absolutely not unique but by the enforcement of a future contract, it would be the basis of a singular relation.

The architecture put forward by the bidders relies on this decomposition, each weighting $W_\beta$ standing for the influence of the architect $\beta$ – one of the $\beta_{max}$ architects involved in the architecture construction – on the design of the solution proposed to the customer or designed for a given market. An example of permutation matrix $\Pi_\beta$ is shown below (only one 1 in any row and only one 1 in any column).

$$\begin{array}{c} \phantom{f_1}\ f_1\ f_2\ f_3\ f_4\ f_5\ f_6\ f_7 \\ \begin{array}{c} f_1 \\ f_2 \\ f_3 \\ f_4 \\ f_5 \\ f_6 \\ f_7 \end{array} \left( \begin{array}{ccccccc} 0 & 1 & 0 & 0 & 0 & 0 & 0 \\ 0 & 0 & 0 & 1 & 0 & 0 & 0 \\ 0 & 0 & 0 & 0 & 0 & 1 & 0 \\ 0 & 0 & 1 & 0 & 0 & 0 & 0 \\ 0 & 0 & 0 & 0 & 0 & 0 & 1 \\ 1 & 0 & 0 & 0 & 0 & 0 & 0 \\ 0 & 0 & 0 & 0 & 1 & 0 & 0 \end{array} \right) = \Pi_\beta \end{array} \tag{36}$$

The order of the activations is the following (according to two cycles $\sigma_\beta^k$'s):

$$\sigma_\beta^1 : f_1 \to f_2 \to f_4 \to f_3 \to f_6 (\to f_1...)$$
$$\sigma_\beta^2 : f_5 \to f_7 (\to f_5...)$$
(37)

All the uncertainty of $M_F$ is transferred to the architects of the solution through the weightings $W_\beta$'s while the permutation matrix of each architect characterizes the certainty to get a defined piece of solution.

Let the costing operator be $E = diag(E_1, E_2, ..., E_{m_F})$ identified through a professional cost estimate in terms of work necessary for the solution to be realized. Here, the work $E$ integrates the unitary, or normalized to 1, time cost. Let the information operator be $I = diag(I_1, I_2, ..., I_{m_F})$ which is the result of analyses of the state of the art for the solution construction. The bid estimate is based on the matrix $B = EM_F I$ and a Rough Order of Magnitude (ROM) is drawn up within an interval which is sometimes defined by the customer and according to the following mechanism, where $I_\theta$ and $E_P$ characterize respectively the best knowledge and the highest professionalism level in the field of the solution:

$$\sum_{j=1}^{m_F} E_i \omega_{i,j} I_j = E_i \sum_{j=1}^{m_F} \omega_{i,j} I_j \geq E_i I_\theta, \quad \forall i$$
$$\sum_{i=1}^{m_F} E_i \omega_{i,j} I_j = I_j \sum_{i=1}^{m_F} E_i \omega_{i,j} \leq I_j E_P, \quad \forall j$$
(38)

and

$$I_\theta Tr(E) \leq ROM \leq E_P Tr(I)$$
(39)

the knowledge $I_\theta$ and the professionalism $E_P$ being such that:

$$I_\theta \leq Tr(I), \quad Tr(E) \leq E_P$$
(40)

These conditions account for the fact that the best knowledge is always challenged by a new project which is an opportunity to generate new concepts and the result of a specific reflection, whereas the highest level of professionalism drives the project above the work [cost] estimate, especially when the management, under the customers' pressure, leads its collaborators to excel themselves.

Applying the $E$ and $I$ operators to each component $\Pi_\beta$ ($\Pi_\beta = \left[\omega_{i,j}^\beta\right]_{1\leq i,j\leq m_F}$) of the decomposition of $M_F$:

$$\forall i \; \exists! k_i \; \sum_{j=1}^{m_F} E_i \omega_{i,j}^\beta I_j = E_i I_{k_i} \geq E_i I_\theta^\beta$$
$$\forall j \; \exists! l_j \; \sum_{i=1}^{m_F} E_i \omega_{i,j}^\beta I_j = E_{l_j} I_j \leq E_P^\beta I_j \tag{41}$$

The $\beta$ estimate [of the necessary budget] $b_\beta$ provided by the corresponding architect satisfies:

$$b_\beta = \sum_{j=1}^{m_F} E_{l_j} I_j = \sum_{i=1}^{m_F} E_i I_{k_i} \tag{42}$$

It is possible to specify mathematically – what is not developed here – a lower bound and an upper bound to all $b_\beta$'s, according to the rearrangement inequality (Hardy et al. 1952) applied to the series of $E$ and $I$ eigenvalues. At the bid stage, only positive eigenvalues are considered.

The singular case $\beta_{\max} = 1$ when $M_F = \Pi_1$ entails $I_\theta^1 \leq \min_i(I_i)$, $E_P^1 \geq \max_i(E_i)$ and eventually:

$$I_\theta^1 Tr(E) \leq b_1 \leq E_P^1 Tr(I) \tag{43}$$

For this singular case the upper bound to $E_P^1$ can be lower than $Tr(E)$ what confirms that a quick agreement between the customer and a preselected solution provider can lead to lower financial charges but without the guarantee of the best competitive professionalism. The lower bound to $I_\theta^1$ is far less than $Tr(I)$ what shows that the selected firm has to create more information than available on the market or than the inputs from the customer.

For the general case, $\beta_{max} > 1$, the lower and the upper bounds to $b_\beta$ are specific and, beyond the combinatorial aspects, imposed by the firms' budget and prospect:

$$b_\beta^- \leq b_\beta \leq b_\beta^+ \tag{44}$$

Thresholds can then be defined:

$$W^+ = \frac{\sum_\beta b_\beta^+ W_\beta}{\sum_\beta b_\beta^+}, \quad W^- = \frac{\sum_\beta b_\beta^- W_\beta}{\sum_\beta b_\beta^-} \tag{45}$$

and a weighted mean is:

$$W^\pm = \frac{\sum_\beta b_\beta W_\beta}{\sum_\beta b_\beta} \tag{46}$$

When $\beta_{max} = 1$, the negotiations between the firm and the customer are led by mutual agreement, whereas in the general case the agreement is the result of a competitive process in which there can be a competition – sometimes a collaborative one – between the architects or designers inside a firm or between tenderers within the framework of an invitation to bid. In some complex situations the customer can also push several complementary firms to cooperate for a given contract whereas other kinds of cooperation are motivated by a market and the related prospects. Whatever the industrial and financial set-up, competitiveness can be summed-up through the following diagram.

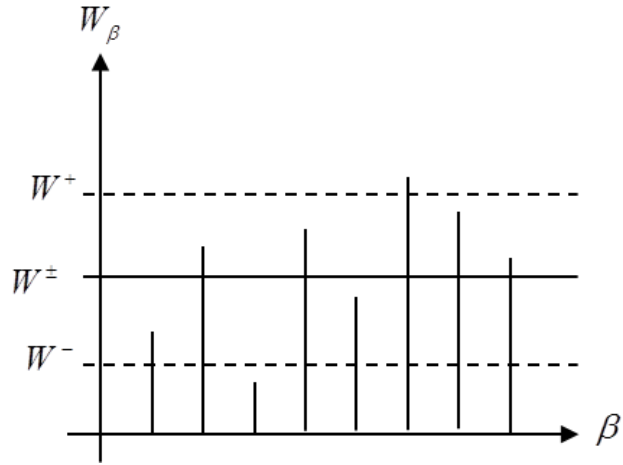

**Fig. 6.** Selection of solution components

The selection of the most influential architect or designer $\left(W_\beta > W^+\right)$ may be detrimental to the profitability of the project particularly if it is associated with high a budget requirement: the greater the value of $b_\beta W_\beta$, the higher the contribution to $W^\pm$ and so the weighted mean is pulled to the top sometimes only by a few designers, if not a single one, who promote expensive solutions. At the opposite, low values of $b_\beta W_\beta$ pull $W^\pm$ towards the bottom and provide low attractiveness solutions.

When $b_\beta$ is low and $W_\beta$ is great, the corresponding solution is optimized and also attractively presented, whereas if $b_\beta$ is high and $W_\beta$ low, this case characterizes a high budget solution put forward by a low credibility designer.

This classification does not provide any certainty for a selection decision but makes clearer the ambiguity in the selection process which relies much on the feeling of making a real bargain when choosing a project manager. But above all such a feeling relies on concrete elements: the reputation of the chosen firm based on the quality of its make and the clearest possible sight on its sustainability or its compliance to standards that is compatible with maintaining the solution at the expected level of use for a time length in adequacy with the customer profitability objectives.

In some circumstances, contenders embody their own standards while they influence the evolution of the state of the art in prescriptive organizations up to government ones and to world authorities. In this case the selection is highly economic, not to say political when customers are institutions such as ministries or governments. The whole economy of a country, or of a group of countries, can then be concerned.

In front of the choice after application of selection criteria and according to the stakes of a market, a customer can decide to select several pieces of solution $\beta$. According to the purpose, calculation and "feeling" of the customer, the choice consists in selecting the right architects or designers $\beta$ who support the corresponding solutions. Logically the influence magnitudes $W_\beta$'s crystallize an assessment of these solutions: excellent $(W_\beta > W^+)$, attractive $(W^+ \geq W_\beta \geq W^-)$ and unsatisfactory $(W^- > W_\beta)$ in the case of homogeneity of the solutions. Out of such a homogeneity, cases where $W^+ < W^-$ can be found characterizing a kind of market pathology.

However, this article does not go beyond this theoretical assessment, the truth in decision-making being that it belongs to decision-makers. In other terms any speculation on a way for a customer to choose a contractor would be a pure intellectual and speculative bubble. So it is supposed that a decision is made and that the choice of the customer falls on a given set B of $\beta's$. Let $\Sigma_\beta = \{\sigma_\beta^k\}_k$ be the set of the cycles associated with the permutation matrix $\Pi_\beta$. The matrix $M_B = \left[ \omega_{i,j}^B \right]$ is then made up with the superposition of all the cycles $\sigma_\beta^k$'s elaborated for all the selected solutions $\beta$'s:

$$\begin{aligned} \omega_{i,j}^B &= 1 \; iff \; \exists \beta \in B \,|\, \omega_{i,j}^\beta = 1 \\ \omega_{i,j}^B &= 0 \; iff \; \forall \beta \in B \; \omega_{i,j}^\beta = 0 \end{aligned} \tag{47}$$

By construction of the ROM and of $M_B$, the industrial set-up required by the customer involves a financial assessment which is still between the initial lower and upper bounds of the ROM (according to $B_B = E M_B I$) but the arrangement imposed by the customer can generate synergies which may reduce costs or lead to extra expenditures for coordination need. So, the cumulative contribution of $\beta$'s is planned as being within the ROM estimate.

*4.2. The Industrialization of a Solution by a Unique Contractor*

Each function $f_i$ uses the set of technical components $X_i = \{\chi_{i,j}\}_j$ according to events $(A_{i,j})$.

The industrialization of the solution $\beta$ is embodied in the solution matrix built from $\Pi_\beta$ where each 1 is replaced by the matrix $T_{i,l}^\beta$ of technical components interaction while each 0 is replaced by the null matrix. The example of $\Pi_\beta$ leads to considering the following doubly stochastic matrix $\Pi_\beta^\uparrow$ in which each block matrix $T_{i,l}^\beta = \left[ t_{k,j}^\beta \right]_{1 \leq k, j \leq d_l^\beta}$ is also bistochastic in order to characterize the probability for each component $\chi_{i,j} \left( j = 1, 2, ..., d_l^\beta \right)$ to activate [or interact with] any other technical component used to perform $f_i$.

$$\Pi_\beta^\uparrow = \begin{pmatrix} & d_1^\beta & d_2^\beta & d_3^\beta & d_4^\beta & d_5^\beta & d_6^\beta & d_7^\beta & \\ 0 & T_{1,2}^\beta & 0 & 0 & 0 & 0 & 0 \\ 0 & 0 & 0 & T_{2,4}^\beta & 0 & 0 & 0 \\ 0 & 0 & 0 & 0 & 0 & T_{3,6}^\beta & 0 \\ 0 & 0 & T_{4,3}^\beta & 0 & 0 & 0 & 0 \\ 0 & 0 & 0 & 0 & 0 & 0 & T_{5,7}^\beta \\ T_{6,1}^\beta & 0 & 0 & 0 & 0 & 0 & 0 \\ 0 & 0 & 0 & 0 & T_{7,5}^\beta & 0 & 0 \end{pmatrix} \begin{matrix} d_2^\beta \\ d_4^\beta \\ d_6^\beta \\ d_3^\beta \\ d_7^\beta \\ d_1^\beta \\ d_5^\beta \end{matrix} \qquad (48)$$

The order $N_\beta$ of $\Pi_\beta^\uparrow$ is the sum of the orders of the $T_{i,l}^\beta$'s: $N_\beta = \sum_{l=1}^{m_F} d_l^\beta$. The initial costing operator $E$ is split according to the vertical partition of the orders while the information operator $I$ is split according to the horizontal partition of the orders (below, the case of the example):

$$E_\beta^\uparrow = diag\left( \underbrace{E_{1,1}^\beta, ..., E_{1,d_2^\beta}^\beta}_{E_1}, \underbrace{E_{2,1}^\beta, ..., E_{2,d_4^\beta}^\beta}_{E_2}, \underbrace{E_{3,1}^\beta, ..., E_{3,d_6^\beta}^\beta}_{E_3}, ... E_{7,d_5^\beta}^\beta \right)$$

$$I_\beta^\uparrow = diag\left( \underbrace{I_{1,1}^\beta, ..., I_{1,d_1^\beta}^\beta}_{I_1}, \underbrace{I_{2,1}^\beta, ..., I_{2,d_2^\beta}^\beta}_{I_2}, ... I_{7,d_7^\beta}^\beta \right) \qquad (49)$$

In the example:

$$E_1 = \sum_{i=1}^{d_2^\beta} E_{1,i}^\beta, \ E_2 = \sum_{i=1}^{d_4^\beta} E_{2,i}^\beta, \ E_3 = \sum_{i=1}^{d_6^\beta} E_{3,i}^\beta, ...$$

$$I_1 = \sum_{i=1}^{d_1^\beta} I_{1,i}^\beta, \ I_2 = \sum_{i=1}^{d_2^\beta} I_{2,i}^\beta, ... \qquad (50)$$

Before considering any profitable margin, the budget determined within the ROM, and then according to the contract, is broken down and the allocation matrix of the contractor $\beta$ is $B_\beta^\uparrow = E_\beta^\uparrow \Pi_\beta^\uparrow I_\beta^\uparrow$. As seen previously, the budget of the sole contractor $\beta$ is clearly much lower than the upper bound of the initial ROM.

As shown in the appendix A, the decomposition of $\Pi_\beta^\uparrow$ into permutation matrices, according to the Birkhoff – von Neumann theorem, is such that each $T_{i,l}^\beta$ is itself decomposed into permutation matrices with the same coefficients as those of the expansion of $\Pi_\beta^\uparrow$:

$$\Pi_\beta^\uparrow = \sum_\alpha W_\alpha \Pi_\alpha^{N_\beta}, \quad \sum_\alpha W_\alpha = 1, W_\alpha \geq 0 \tag{51}$$

Still according to the example ($\pi_{i,l}^\alpha$ is a permutation matrix of order $d_l^\beta$):

$$\Pi_\alpha^{N_\beta} = \begin{pmatrix} 0 & \pi_{1,2}^\alpha & 0 & 0 & 0 & 0 & 0 \\ 0 & 0 & 0 & \pi_{2,4}^\alpha & 0 & 0 & 0 \\ 0 & 0 & 0 & 0 & 0 & \pi_{3,6}^\alpha & 0 \\ 0 & 0 & \pi_{4,3}^\alpha & 0 & 0 & 0 & 0 \\ 0 & 0 & 0 & 0 & 0 & 0 & \pi_{5,7}^\alpha \\ \pi_{6,1}^\alpha & 0 & 0 & 0 & 0 & 0 & 0 \\ 0 & 0 & 0 & 0 & \pi_{7,5}^\alpha & 0 & 0 \end{pmatrix} \tag{52}$$

and

$$T_{i,l}^\beta = \sum_\alpha W_\alpha \pi_{i,l}^\alpha \tag{53}$$

Applying the operators $E_\beta^\uparrow$ and $I_\beta^\uparrow$ to $\Pi_\alpha^{N_\beta}$ leads to considering the application of their partitions to the corresponding $\pi_{i,l}^\alpha$'s ($\pi_{i,l}^\alpha = \left[\omega_{k,j}^\alpha\right]_{1 \leq k,j \leq d_l^\beta}$):

$$\begin{aligned} E_{\beta i}^\uparrow &= diag\left(E_{i,1}^\beta, E_{i,2}^\beta, ..., E_{i,d_l^\beta}^\beta\right); E_i = Tr\left(E_{\beta i}^\uparrow\right) \\ I_{\beta l}^\uparrow &= diag\left(I_{l,1}^\beta, I_{l,2}^\beta, ..., I_{l,d_l^\beta}^\beta\right); I_l = Tr\left(I_{\beta l}^\uparrow\right) \end{aligned} \tag{54}$$

$$\forall k \; \exists! \kappa_k \; \sum_{j=1}^{d_l^\beta} E_{i,k}^\beta \omega_{k,j}^\alpha I_{l,j}^\beta = E_{i,k}^\beta I_{l,\kappa_k}^\beta$$

$$\forall j \; \exists! \lambda_j \; \sum_{k=1}^{d_l^\beta} E_{i,k}^\beta \omega_{k,j}^\alpha I_{l,j}^\beta = E_{i,\lambda_j}^\beta I_{l,j}^\beta$$

(55)

The $\alpha$ estimate of the budget $a_\alpha^\beta$ of the department $D_\alpha$ for its contribution to the production of the function $g_{i'}$ (under the design of $f_i$) is then:

$$a_\alpha^\beta = \sum_{k=1}^{d_l^\beta} E_{i,k}^\beta I_{l,\kappa_k}^\beta = \sum_{j=1}^{d_l^\beta} E_{i,\lambda_j}^\beta I_{l,j}^\beta \qquad (56)$$

The budget associated with $T_{i,l}^\beta$ is $E_i I_l$ in order for the departments $D_\alpha$'s to build the function $f_i$. This anticipated budget satisfies (according to the initial costing of $f_i$):

$$E_i I_l = Tr\left(E_{\beta i}^\uparrow\right) Tr\left(I_{\beta l}^\uparrow\right) \qquad (57)$$

Through the same mechanism as the one used for the selection of the participants in the industrial arrangement concerning $M_F$, each function $f_i$ design is an opportunity for optimization through the selection of the relations between the technical components that enable to produce $g_i$. So according to the technical policy of the firm which promotes some of its products or developments (the set A gathers the $\alpha$'s corresponding to this promotion), $T_{i,l}^\beta$ leads to $T_{i,l}^A = \left[t_{k,j}^A\right]$ such that:

$$\begin{aligned} t_{k,j}^A &= 1 \; iff \; \exists \alpha \in A \,|\, \omega_{k,j}^\alpha = 1 \\ t_{k,j}^A &= 0 \; iff \; \forall \alpha \in A \; \omega_{k,j}^\alpha = 0 \end{aligned} \qquad (58)$$

Clearly, $T_{i,l}^{A}$ involves a solution the budget of which is lower than the anticipated one for $T_{i,l}^{\beta}$. And it is so for any block $T_{i,l}^{\beta}$ of $\Pi_{\beta}^{\uparrow}$.

Such reasoning applies to any potential contributor $\beta$ through the enhancement of $\Pi_{\beta}$ and within the initial budget for building the set of the functions expected by the customer but with the functional relations specific to the contributor's design, to its organization and to the way how it delivers the solution to the customer. Rigorously, the structure of this organization must be adapted according to the decomposition of $\Pi_{\beta}^{\uparrow}$ (and of all the $T_{i,l}^{\beta}$) into permutation matrices, what can be the case for a big contract of a firm to provide one of its main customer with high stake services or [mass-produced] products or also when the firm has designed a good which requires a specific organizational structure for its development and its production to the market.

### 4.3. The Industrialization of a Solution Within an Industrial Arrangement

The matrix $M_F$ gives concrete expression to the negotiations between the customer and the firms or the consortiums that are bidding for providing the customer with a solution. The expectations in terms of functional relations and the answers by the firms can be synthetized in the following condition:

$$\omega_{i,j} \neq 0 \Leftrightarrow \exists \beta \mid \omega_{i,j}^{\beta} = 1 \tag{59}$$

As seen previously, the selection of those firms leads to turning the uncertainties of $M_F$ – associated with the customer's expectations – into industrial achievement while leaving outside the perimeter of the solution designed by the firms, which have been chosen, a part of the expected functional relations:

$$\exists \omega_{i,j} \neq 0 \mid \omega_{i,j}^{B} = 0 \tag{60}$$

Let $B_{i,j}$ be the following set related to the process of selection:

$$B_{i,j} = \left\{ \beta \in B \mid \omega_{i,j}^{B} = 1 \text{ and } \omega_{i,j}^{\beta} = 1 \right\} \tag{61}$$

Logically for a straightforward profitability, the comparative advantage principle involves choosing $T_{i,j}^{\beta}$, then $T_{i,j}^{A(\beta)} = \left[ t_{k,l}^{A(\beta)} \right]_{1 \leq k,l \leq d_{j}^{\beta}}$ and so $\beta = \beta_{i,j}^{c}$ such that:

$$H_{i,j}\left(\beta_{i,j}^{c}\right) = \min_{\beta \in B_{i,j}} \left\{ H_{i,j}(\beta) \right\}, \quad H_{i,j}(\beta) = \sum_{k=1}^{d_{j}^{\beta}} \sum_{l=1}^{d_{j}^{\beta}} E_{i,k}^{\beta} t_{k,l}^{A(\beta)} I_{j,l}^{\beta} \tag{62}$$

But this choice, although providing an element of solution that seems competitive and cost-effective, does not ensure the unicity of $\beta_{i,j}^{c}$ within the consortium. In fact if the set of $\beta_{i,j}^{c}$ is not a singleton, the firms that cooperate in the project have to decide which one of the $\beta_{i,j}^{c}$'s will actually be chosen.

On the one hand, this decision is typically a matter of industrial governance and is based on the comparative know-how of teams of the firms involved in the customer satisfaction. At that stage the governance determination is motivated by ambition, the search of success and acknowledgement at the highest corporate level and this determination dispels the doubts that could subsist or arise as regards the outcome of the project or the continuity of the relationships between the participants in the industrial arrangement or with their customer. At the stage of the bid process, the operators $E$ and $I$ have only positive eigenvalues, thus each term of energy, work or adaptive utility $E_i$ and each associated piece of information $I_j$ are supposed positive at that time. And this positivity is maintained all along the transition between the bid and the project by the fact that the contract is signed [by all the stakeholders] and by the governance determination which transfers uncertainty to the technical teams in charge of the project achievement.

But on the other hand, when it is about building the real solution within the contract framework, the industrial constraints are such that the decision for the selection must put forward the element of solution $T_{i,j}^{\beta}$ which is sustainable and yielding the least possible negative externalities.

Up to here the budget of the project is under control and proves to be profitable from every point of view.

The decompositions of $E_i$ and $I_j$ are known for the exploratory element of solution as:

$$E_i = \sum_{k=1}^{d_j^{\beta}} E_{i,k}^{\beta}, \quad I_j = \sum_{l=1}^{d_j^{\beta}} I_{j,l}^{\beta} \tag{63}$$

In this exploration it appears [on the more or less long term] that terms of these decompositions can be negative in such a way that $H_{i,j}(\beta)$ could also be negative although $E_i$ and $I_j$ are positive. An easy analysis can establish that externalities are quantified through $H_{i,j}(\beta)$. At the level of a solution element $T_{i,j}^{\beta}$, a minor default, $\exists k,l\ E_{i,k}^{\beta} t_{k,l}^{A(\beta)} I_{j,l}^{\beta} < 0$ and $H_{i,j}(\beta) > 0$, will not necessarily cause a noticeable malfunction with a high impact on the global solution. Such technical issues, when they are known, can be overcome by a "workaround" without preventing the solution from being used normally. So, the potential externality lies in the fact that this minor default involves a little more work to apply the workaround when needed during the running of the solution. A major default $H_{i,j}(\beta) < 0$ is a real cause of problem the consequence of which is a malfunction with an impact on the solution that necessitates developing operational measures and a specific communication towards the customer so as to mitigate his or her dissatisfaction. This case does not constitute a clear externality because the service or good is delivered although not according to the contractual expectations nor carried professionally.

Although it is doubtful that information could have a negative value, it is possible to specify that ignorance corresponds to $I_{j,l}^{\beta} = 0$, efficient knowledge to $I_{j,l}^{\beta} > 0$ and paradoxical thinking to $I_{j,l}^{\beta} < 0$. Whereas a technical failure can clearly be due to a negligence, $E_{i,k}^{\beta} t_{k,l}^{A(\beta)} I_{j,l}^{\beta} < 0$ and $I_{j,l}^{\beta} > 0$, it is conceivable that such a failure can be corrected using counterintuitive information: $E_{i,k}^{\beta} t_{k,l}^{A(\beta)} I_{j,l}^{\beta} > 0$ with $I_{j,l}^{\beta} < 0$. Furthermore sometimes technical components can be used out of their initial range of functioning ($E_{i,k}^{\beta} < 0$) and so profitably when they are twisted $\left(I_{j,l}^{\beta} < 0\right)$ in order to create uncommon functions. On the other hand malicious thinking can be characterized by $E_{i,k}^{\beta} t_{k,l}^{A(\beta)} I_{j,l}^{\beta} < 0$ with $I_{j,l}^{\beta} < 0$: the component is used for something it was not meant for.

The production-consumption structure analysis would lead to specify the $E_i^R$'s according to entangled cycles of $T_{i,j}^{A(\beta)}$ and to the corresponding terms of $H_{i,j}(\beta)$, enabling to find the final nature of $g_i$'s. But this combinatorial work is strongly dependent on the final user of the produced system. It is thus possible to distinguish a production value for the project, as shown in equation (33), from a value to the consumer: the way how a product is built anticipates a wide range of uses but whatever the initial purpose, the consumption relies on the use customization. Both values are taken into account in the equations of the market dynamics of the next section.

Finally and roughly when $E_{i,k}^{\beta} < 0$ for all k's while using efficient knowledge, $E_i^R$ is not delivered at the corresponding ROM anticipated level $E_i$ and the default has a high impact on the customer's business and then on his own customers, what is an explicit negative externality. In order to prevent the solution from such externalities, as in the ROM, framing boundaries can be imposed to $H_{i,j}(\beta)$ as an internal contract, between the project management and the technical teams, relying on the requirements coming from the customer and structured into technical ones by the designers or architects of the solution.

But in an always changing world, the evolutions of standards or of practices and skills are the main reason why solution failures occur while running a complex system: this is commonly what is called obsolescence. In the present case the variety of the firms and their relative responsibilities in the production of the solution elements lead to examine interfaces between the various contributions in order to express needs of interoperability between those elements each of which obeys a different obsolescence cycle. Treating the obsolescence of one given component or element can have serious impacts on others and then on the global solution with possible failures. So a function provided according to the state of the art at a given time ($E_i^R > 0$) may prove to be a negative externality ($E_i^R < 0$) after the work of time: because of standards and regulations changes or outdated components when skills for their maintenance are no longer available. The conditions of the obsolescence handling are generally included in the contractual requirements, especially when the solution running is expected on a long term.

Whatever the solution characterized by $M_B$ each chosen element $T_{i,j}^{\beta}$ has its own order $d_j^{\beta}$. So building $M_B^{\uparrow}$, by the replacement of each 1 in $M_B$ by a bistochastic matrix $T_{i,j}^{\beta}$ and each 0 by a null block matrix, what involves the uniformity of their orders, is generally not possible. Such uniformity could be a first step towards interoperability between the elements of solution coming from the teams of the industrial arrangement different firms. However even through uniform orders, $M_B^\uparrow$ is not doubly stochastic, except if it concerns a sole firm which gained the contract with the customer, what refers to $\Pi_\beta^\uparrow$. Furthermore the theorem of Birkhoff – von Neumann does not apply to $M_B^\uparrow$ so there is no rule for its possible decomposition and there is no constraint on the internal organization of each firm of the industrial arrangement in spite of the macro-organization of the project management through the set of the $\Pi_\beta$'s and the selection of $\beta$'s.

The industrialization of a solution within an industrial arrangement relies formally on the commitment of each member of the chosen consortium in accordance with the model of a sole contractor. The development of this paragraph emphasizes the problematic of the externalities that applies to the unique firm which may also have interoperability constraints to satisfy at a lesser level and obsolescence to manage. In any case the chain of uncertainty can be followed from the ROM estimate to the technical contingencies and it seems clear that any industrial contract is a real opportunity and that profitability depends on the commitment of each actor of this chain particularly within a matrix organization.

## 5. The fundamental equations of the market dynamics

The equation linking the payoff to the value can be set down within a supply policy:

$$\kappa_{Os} \frac{dP_s}{dm_s} = P_s - V_s \tag{64}$$

The increase in $P_s$ is due to $P_s$ and to the delivery of the value $V_s$. $\kappa_{Os}$ is a constant homogeneous to a complexity – or reduced capacity – that could be positive or negative and which is a reference in the chosen frame.

### 5.1. Pure Speculative Zero Value Dynamics

When there is no creation of value in the chosen frame, $V_s = 0$ and:

$$P_s = P_{Os} \exp \frac{m_s - m_{Os}}{\kappa_{Os}} \tag{65}$$

a) If $\kappa_{Os} < 0$

For the supply, $C_s$ and $m_s$ are positive according to figure 2, and when $P_{Os} > 0$, the evolution of $P_s$ can be interpreted as an impoverishment of the firm or agent who chose a strategy with no industrial value foundations: the farther the strategy construction goes – $m_s$ increases –, the lower the payoff. This is the case when theoreticians design a system that cannot be built within a reasonable "time to market" and so the designed product will not reach the market. Or if it does, when the investment is sufficient, it might be for the profit of the firms which are financially powerful enough to invest on this product's development.

If $P_{Os}$ is negative relatively to the frame of reference Os, $P_s$ increases but stays negative. The reason for this negativity is not due to the capability of the members of the organization who deliver the product strategy corresponding to the payoff. But it comes from their work $(E_i\text{'s})$ that constitutes a negative externality for the global economy. An extreme example of such an externality is terrorism the payoff of which stays negative, whatever the functions used.

b) If $\kappa_{Os} > 0$

$C_s$ and $m_s$ being positive (figure 2), the speculation is a success whenever it is based on a frame where the work can be considered positively: even intellectual positions and design thinking may lead to positive externalities and then to an increasing payoff when concepts or strategies are developed.

Conversely, negative externalities are all the more important since the chosen frame determines the [initial] reference work as harmful to the global economy and that the intention of harming is reinforced by the increasing number of functions used to this end. This emphasizes the risks of growth of every kind of terrorism but also of any escalation of violence, even in an economic field.

*5.2. The Equilibrium of the Theory*

The equilibrium is reached when:

$$\kappa_{Os} \frac{dP_s}{dm_s} = P_s - V_s = 0 \Leftrightarrow P_s = V_s \qquad (66)$$

As established by Nash (1951) in the framework of the non-cooperative game theory, value equates payoff in some specific conditions of equilibrium. Chauvet (2013, 2015) used this result in the limited context of the basic economic law of supply and demand.

The equilibrium equation is equivalent to:

$$\frac{1}{C_s} \sum_{i=1}^{i=\lceil m_s \rceil} E_i^S = \frac{1}{c_s} \sum_{i=1}^{i=m_R} E_i^R \rho \qquad (67)$$

When the designed functions are exactly the performed ones, this equation becomes:

$$\rho C_s = c_s \Leftrightarrow \rho \frac{dm_s}{d\rho} = c_s \qquad (68)$$

That is the equation which characterizes the work utility and leads to the graphic expression of the law of supply and demand as seen previously (figure 4) and according to Chauvet (2013).

When $F_s$ and $F_R$ are not totally equivalent, that basic law is slightly modified to take into consideration the maturity $M_s$ of the valued product to the market (Chauvet 2015).

Furthermore, introducing this maturity in the calculation of the work utility gives:

$$W_{us} = \rho \frac{dm_s}{d\rho} = \frac{c_s}{M_s} \tag{69}$$

The introduction of the maturities for [the demand and] the supply has a serious impact on the [mutual] growth of the market but the calculation of $\Delta K$ shows a tricky behavior from which no simple conclusion emerges. This subject is postponed to the end of this article which deals with a sustainable growth pattern.

### 5.3. Out of Equilibrium

When replacing $P_s$ and $V_s$ by their previously defined expressions, the fundamental equation of the market dynamics becomes:

$$\kappa_{Os} \frac{d}{dm_s}\left[\frac{d\rho}{dm_s}\sum_{i=1}^{i=\lceil m_s \rceil} E_i^s\right] = \frac{d\rho}{dm_s}\sum_{i=1}^{i=\lceil m_s \rceil} E_i^s - \frac{1}{c_s}\sum_{i=1}^{i=m_R} E_i^R \rho \tag{70}$$

which is also, most generally and according to the appendix B:

$$\kappa_{Os}\frac{d^2\rho}{dm_s^2} - \frac{d\rho}{dm_s} + \frac{M_s}{c_s}\rho = 0\,;\, M_s = \frac{\sum_{i=1}^{i=m_R} E_i^R}{\sum_{i=1}^{i=\lceil m_s \rceil} E_i^s} \tag{71}$$

The characteristic equation of that differential equation is:

$$\kappa_{Os}\gamma^2 - \gamma + \frac{M_s}{c_s} = 0 \tag{72}$$

And for $\kappa_{Os} > 0$ its solutions are:

$$\gamma_\pm^s = \frac{1 \pm \sqrt{1 - 4\frac{\kappa_{Os} M_s}{c_s}}}{2\kappa_{Os}} \tag{73}$$

Finally, the solution of the differential equation is:

$$\rho_s^>(m_s) = \rho_s^+ \exp(\gamma_+^s m_s) + \rho_s^- \exp(\gamma_-^s m_s) \tag{74}$$

This solution is also written:

$$\rho_s^>(m) = (\rho_s^+, \gamma_+^s) + (\rho_s^-, \gamma_-^s), \, m = m_s \geq 0 \tag{75}$$

If $\kappa_{Os} < 0$:

$$|\kappa_{Os}|\gamma^2 + \gamma - \frac{M_s}{c_s} = 0 \tag{76}$$

and

$$\gamma_{|\pm|}^s = \frac{-1 \pm \sqrt{1 + 4\frac{|\kappa_{Os}| M_s}{c_s}}}{2|\kappa_{Os}|} \tag{77}$$

The solution is then:

$$\rho_s^<(m) = (\rho_s^{|+|}, \gamma_{|+|}^s) + (\rho_s^{|-|}, \gamma_{|-|}^s), \, m = m_s \geq 0 \tag{78}$$

*5.4. Dual Approach Within a Demand Policy*

From the viewpoint of a demand policy, the fundamental equation of market dynamics is:

$$\kappa_{Od} \frac{dP_d}{dm_d} = P_d + V_d \qquad (79)$$

The increase in $P_d$ is due to $P_d$ and to the gain of the value $V_d$.

When $V_d = 0$, the solution of this equation is:

$$P_d = P_{Od} \exp \frac{m_d - m_{Od}}{\kappa_{Od}} \qquad (80)$$

a) If $\kappa_{Od} < 0$

For the demand, $C_d$ and $m_d$ are negative according to figure 2, the investment always goes with a negative payoff, apart from when this demand is made for negative externalities consequences, for instance, of a polluting industry (negative work) needing other industrial organizations [demand] for cleaning up. If $P_{Od}$ is positive, in the "Od" frame, that also implies a demand for negative work and $P_d$ rises when $m_d$ decreases, what can be understood as a well-paid getting rid of pollution.

b) If $\kappa_{Od} > 0$

$C_d$ and $m_d$ being negative, the demand policy confirms that the efforts spent on the market are generally associated with a negative – but also increasing – payoff. Here $P_d$ increases with the number of functions appealed to when $P_{Od}$ is negative: the higher the scale of the demand, the greater the corresponding payoff. But $P_d$, positive for a negative work, decreases with $m_d = -|m_d|$ when the corresponding externalities – pollution, harmful products or behaviors, etc. – are taken in charge by the demand as engagement to clean up. The more is done for a sustainable economy, the less the reward to the benefactor. This is the reason why a relevant regulation must be enforced by the political powers before any irreparable damage to the environment, in order to avoid the creation of negative externalities and the despondency of those who fight against them, when they appear.

The demand policy and the supply policy are complementary in a speculative thinking, although they lead sometimes to conclusions that could be understood as an endless debate.

At the equilibrium, when the payoff is constant and equates the corresponding non-nil opposite value, the demand policy ends up in the same graphical presentation of the basic law as the supply policy (figure 4), with subtleties already developed by Chauvet (2013).

Out of equilibrium, the solution of the equation of the market dynamics for the demand is, when $\kappa_{Od} > 0$:

$$\rho_d^>(m) = \left(\rho_d^+, -\gamma_+^d\right) + \left(\rho_d^-, -\gamma_-^d\right), m = -m_d \geq 0 \tag{81}$$

with:

$$\gamma_\pm^d = \frac{1 \pm \sqrt{1 + 4\frac{\kappa_{Od} M_d}{c_d}}}{2\kappa_{Od}} \tag{82}$$

If $\kappa_{Od} < 0$:

$$\rho_d^<(m) = \left(\rho_d^{|+|}, -\gamma_{|+|}^d\right) + \left(\rho_d^{|-|}, -\gamma_{|-|}^d\right), m = -m_d \geq 0 \tag{83}$$

with:

$$\gamma_{|\pm|}^d = \frac{-1 \pm \sqrt{1 - 4\frac{|\kappa_{Od}|M_d}{c_d}}}{2|\kappa_{Od}|} \tag{84}$$

*5.5. "Canonical" solutions of the market dynamics equations*

The "canonical" solutions are chosen such that:

$$\begin{aligned} \rho_s^+ = \rho_s^- = \rho_d^+ = \rho_d^- = 1 \\ \rho_s^{|+|} = \rho_s^{|-|} = \rho_d^{|+|} = \rho_d^{|-|} = 1 \end{aligned} \tag{85}$$

and

$$\frac{|\kappa_{Os} M_s|}{c_s} = \frac{|\kappa_{Od} M_d|}{c_d} = 1 \tag{86}$$

with also

$$\kappa_{Os} = |\kappa_{Os}| = \kappa_{Od} = |\kappa_{Od}| = \kappa \tag{87}$$

and the notations: when $\kappa_{Os} > 0$, $\kappa_{Os}$ stands for $\kappa_{Os}$ and when $\kappa_{Os} < 0$, $|\kappa_{Os}|$ stands for $-\kappa_{Os}$.

A perfect fit means that the supply curve is exactly the corresponding demand curve what corresponds to real exchanges between buyers and sellers with high probabilities whereas no intersection between supply and demand curves means that the exchange is not probable.

The figure 7 presents the curves of the canonical solutions according to the following characterization.

The perfect fit of the supply and demand curves is obtained for the real solutions (a) and (b):

a) For $\rho_s^<(m) = \rho_d^>(m)$ shown as the black dots curve for real solutions with

$$\frac{|\kappa_{Os}|M_s}{c_s}=1, \frac{\kappa_{Od}M_d}{c_d}=1 \text{ i.e. } M_s, M_d > 0 \tag{88}$$

$$\rho(m) = \exp\left(-\frac{1+\sqrt{5}}{2}\frac{m}{\kappa}\right) + \exp\left(\frac{-1+\sqrt{5}}{2}\frac{m}{\kappa}\right) \tag{89}$$

b) For $\rho_s^>(m) = \rho_d^<(m)$ shown as the grey dots curve for real solutions with

$$\frac{\kappa_{Os}M_s}{c_s}=-1, \frac{|\kappa_{Od}|M_d}{c_d}=-1 \text{ i.e. } M_s, M_d < 0 \tag{90}$$

$$\rho(m) = \exp\left(\frac{1+\sqrt{5}}{2}\frac{m}{\kappa}\right) + \exp\left(\frac{1-\sqrt{5}}{2}\frac{m}{\kappa}\right) \tag{91}$$

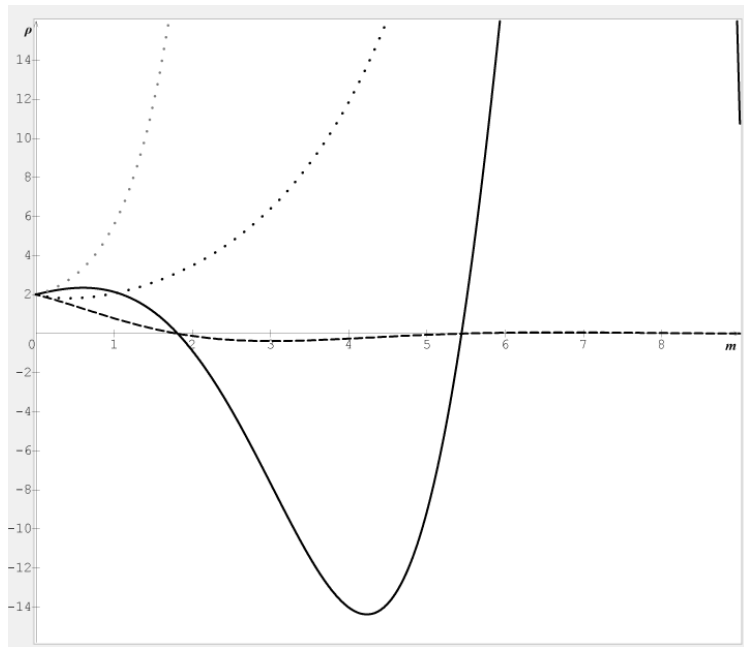

**Fig. 7.** Curves of the canonical solutions $(\kappa = 1)$

The perfect fit of the supply and demand curves is obtained for the complex solutions (c) and (d):

c) For $\rho_s^>(m) = \rho_d^<(m)$ shown as the continuous curve for complex solutions with

$$\frac{\kappa_{Os} M_s}{c_s} = 1, \frac{|\kappa_{Od}| M_d}{c_d} = 1 \; i.e. \; M_s, M_d > 0 \tag{92}$$

$$\rho(m) = 2\exp\left(\frac{1}{2}\frac{m}{\kappa}\right)\cos\left(\frac{\sqrt{3}}{2}\frac{m}{\kappa}\right) \tag{93}$$

d) For $\rho_s^<(m) = \rho_d^>(m)$ shown as the dashes curve for complex solutions with

$$\frac{|\kappa_{Os}| M_s}{c_s} = -1, \frac{\kappa_{Od} M_d}{c_d} = -1 \; i.e. \; M_s, M_d < 0 \tag{94}$$

$$\rho(m) = 2\exp\left(-\frac{1}{2}\frac{m}{\kappa}\right)\cos\left(\frac{\sqrt{3}}{2}\frac{m}{\kappa}\right) \tag{95}$$

Let $\varphi$ be the gold number: $\varphi = \frac{1+\sqrt{5}}{2}$ such that $\varphi^{-1} = \frac{-1+\sqrt{5}}{2}$.

In the real canonical solutions where $M_s, M_d < 0$, the predominant term is $\exp(\varphi \kappa^{-1} m)$, while in the real canonical solutions where $M_s, M_d > 0$, the term which imposes the behavior of the solution is $\exp(\varphi^{-1} \kappa^{-1} m)$. In this second case, there can be an exchange according to the intersection with the complex solution corresponding to the same sign as $M_s$ or $M_d$. But this exchange represents a very low volume on the market. Nevertheless $\varphi^{-1}$ characterizes an opening to wider exchanges than $\varphi$.

Thus, out of the perfect fit there is no significant set of points of exchange between the supply and the demand except for a nil price or time cost (complex solutions) when there is an asymmetry of maturity what corresponds to gifts or thefts or frauds.

It is important to notice that complex solutions with maturity symmetry must only be considered within dynamics of markets. Therefore, oscillations do not provide any information about where and when equilibrium occurs but simply show that, if there is an exchange between supply and demand, it should be found on the complex solution curves.

A first conclusion about the market dynamics is that canonical exchanges between supply and demand can mainly take place, theoretically, when the maturities of the provider and of the buyer are of the same sign. But out of the canonical cases, the behavior of the complexities – or reduced capacities – $(c_s, \kappa_{Os})$ or $(c_s, |\kappa_{Os}|)$ for the supply, and the equivalent for the demand, prevents from knowing to what extent the ranges of the maturities can be decisive in an exchange.

## 6. The probability weighting operator

Decision-making under risk and uncertainty has long been studied from the viewpoint of games and gambling. Generally, choices are presented in order to fit the formalisms worked out to account for rationality. For several decades, the expected utility theory had been the mainstream model available to theoreticians for the description of decision making until the 80's, although Allais (1953) paved the way for non-linearity – in relation to probabilities – of expected utility, through his criticisms of the axioms of the American School of economics and the demonstration of a famous paradox. Tversky and Kahneman (1992) proposed, as a solution to this paradox, a tantalizing and successful theory built on their respective and common works ranging from Prospect Theory to, more specifically, rationality of choice, framing effects, loss aversion, etc., among works of other contributors of this field thanks to whom they found their own theoretical corpus and express their need of experiments.

But as time passes, whatever the theory, it deserves a few touches in order to apply to a wider set of phenomena and to benefit from cross-fertilization with other theories. The Prospect Theory reports studies about gains and losses in a monetary and abstract way in spite of a rigorous "experimental" protocol involving for instance money managers or graduate students from Berkeley and Stanford Universities: "gambling" still seems its only solution to tackle a basic economic problem. Nevertheless the law of supply and demand can be called into question in the light of this theory, any purchase and sale being also a matter of gain and loss: the purchase is a bet on the usefulness of a good or service for a reasonable duration whereas supply is a bet on the quality of a product which implies anticipating its utility. The gain to the buyer is a use at various levels of need, according to the theory of human motivation of Maslow (1943) and her potential loss is any reason for her dissatisfaction. The gain to the "salesman" is essentially monetary and thus, a means of investing in new goods or services production; whereas the loss is a stopping of this virtuous circle. Motivation and complexity are deeply involved in the exchange (Chauvet 2013) and the success on a market depends also on maturities (Chauvet 2015). The parameters that concentrate this information – complexity and maturity – are $\gamma$'s, the solutions of the characteristic equations of the market dynamics differential equations.

*6.1. WEIRD Economics*

According to Henrich et al. (2010), the behavioral and psychological conclusions of studies practiced on people from Western, Educated, Industrialized, Rich, and Democratic (WEIRD) societies, can be questioned as far as a generalization is attempted. Thus, in the same way as Allais criticized the expected utility theory, the "universality" of Prospect Theory can be consolidated. The graduate students who went through the Tversky's and Kahneman's main experiment (1992) where all coming from two great universities of California, sharing the same kinds of higher education, motivation, capabilities of complexity [problems] solving and trained to behave on high maturity markets but also endowed with a mind opened to punctual exchanges with complex market partners. All these features are gathered in $\gamma = \varphi^{-1}\kappa^{-1}$ when $M_s, M_d > 0$ (real canonical solutions).

The probability weighting operator discovered by Tversky and Kahneman (1992) is:

$$w_\gamma(p) = \frac{p^\gamma}{\left(p^\gamma + (1-p)^\gamma\right)^{1/\gamma}} \tag{96}$$

This operator is originally designed with $\gamma = 0.61$ for the weights of positive outcomes and with $\gamma = 0.69$ for the weights of negative outcomes. On the one hand, these values of $\gamma$ have been obtained with only one global experiment and it can be bet that on a larger scale of experimentation there would be a convergence towards $\gamma \cong 0.618 = \varphi^{-1}$. On the other hand it seems relevant to consider that $\kappa$, as a [reduced] capacity of the reference frame, is greater for positive outcomes than for negative ones as it has been observed for $c$ while calculating the value $v$ of an outcome within a project of the firm (figure 5).

The smooth curves obtained by Tversky and Kahneman for both values of $\gamma$ mentioned above can be interpreted as weighting functions, "assuming a linear value function" $v$, both kinds of functions being used in the calculation of the subjective utility of the cumulative prospect theory. So it can be considered that these weighting functions are subject to the influence of $\kappa$ which can create a deviation from the linearity of $w_\gamma(p)$ like $c$ is involved in the non-linearity of $v$. Finally $\kappa = \varphi^{-1}$ is a means of eliminating the probability weighting bias in order to obtain $\gamma = 1$ and imposes a constraint on the maturities and the [reduced] capacities of the canonical real solutions for which $M_s, M_d > 0$:

$$\frac{M_s}{c_s} = \frac{M_d}{c_d} = \kappa^{-1} = \varphi \tag{97}$$

In other terms, the fourfold pattern of risk attitudes showed by Tversky and Kahneman, when it is put into the context of the law of supply and demand, relies on a bias involving maturities and capacities of the actors of the exchange, those actors being firms' representatives or decision-makers or more generally economic agents. Furthermore, this pattern applies essentially to members of WEIRD societies.

At first the data had been analyzed through the ratio of the certainty equivalent of the prospect to the nonzero outcome in monetary terms and the curves had been plotted showing the evolution of this ratio for different values of the corresponding probabilities. In the theory developed in the present article, the monetary aspect is taken into account by $\rho$ while the certainty is paradoxically the inverse of the capability of the agents. So, the ratio initially used by Tversky and Kahneman can be expressed in:

$$\frac{1}{\rho}\left|\frac{d\rho}{dm}\right| \cong w_\gamma(p) \tag{98}$$

This ratio characterizes the bias which distorts the probability weighting by decision-makers as verified experimentally, $p$ and $w_\gamma(p)$ being found between 0 and 1. It also shows homogeneity: for any constant $k > 0$, the transformation $\rho \to k\rho$ keeps it unchanged.

Obviously, the reported experiment relies on the equilibrium for the simulated transactions to be possible as supposed through the perfect fit of the real canonical solutions for positive maturities.

According to the work utility equation and to the condition of cancellation of the probability weighting bias $(\gamma = 1)$ :

$$\frac{1}{\rho}\left|\frac{d\rho}{dm}\right| = \frac{M}{c} = \varphi \tag{99}$$

As $\varphi > 1 = p_{max}$ a logical conclusion is that the fourfold pattern, as a bias, cannot really be eliminated, but it can be exploited to find a way towards reality as it is lived in WEIRD societies. A simplified version of this pattern may help describe what constitutes a business cycle within the framework of markets and eventually it can lead to considering new criteria of well-being.

The figure 8 is obtained for $\gamma = \varphi^{-1}$ and presents the converging series $p_{n+1} = w_\gamma(p_n)$ for $0 < p_0 < p^*$ where $p^*$ satisfies $p^* = w_\gamma(p^*), 0 < p^* < 1$.

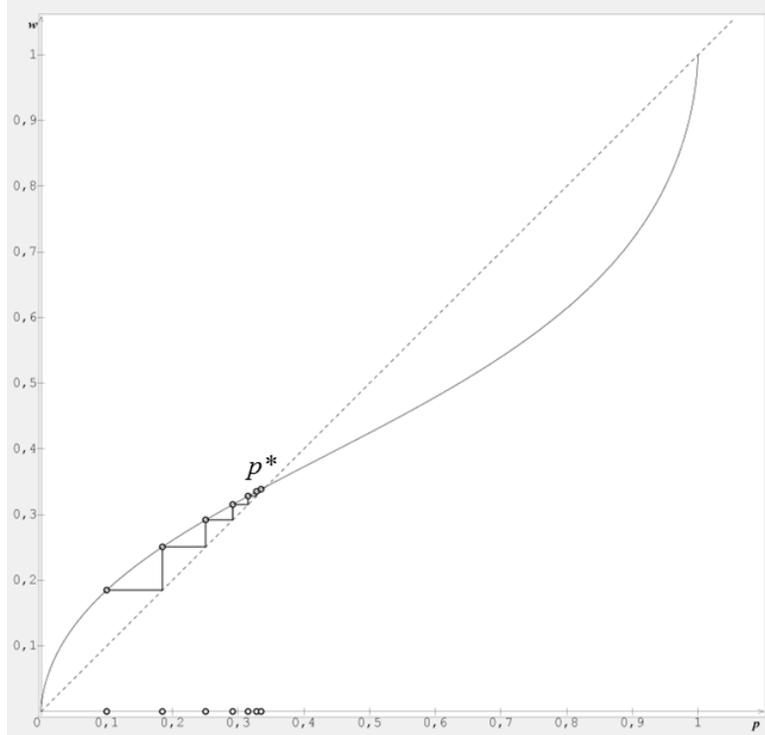

**Fig. 8.** $p_{n+1} = w_\gamma(p_n)$ for $\gamma = \varphi^{-1}$ and $0 < p_0 < p^*$

This figure characterizes the business spirit which reigns in WEIRD societies: as low as $p_0$ can be, the corresponding assessment $w_\gamma(p_0) > p_0$ shows that the motivation of the agent leads him or her to overestimate the probability of an event and for instance to expect a gain for becoming involved in a market or a business. This logic of success is maintained at each step n up to $p^*$ at which point the agent begins feeling trapped in his or her situation what motivates him to change market strategy, business or employment. For those who are persevering and consistent with contractual commitments, this point constitutes the transition from necessary illusions, essential for beginning and building a business, to the real challenge of doubts facing an agent when this business has to be strengthened through a real struggle for a position or a market share.

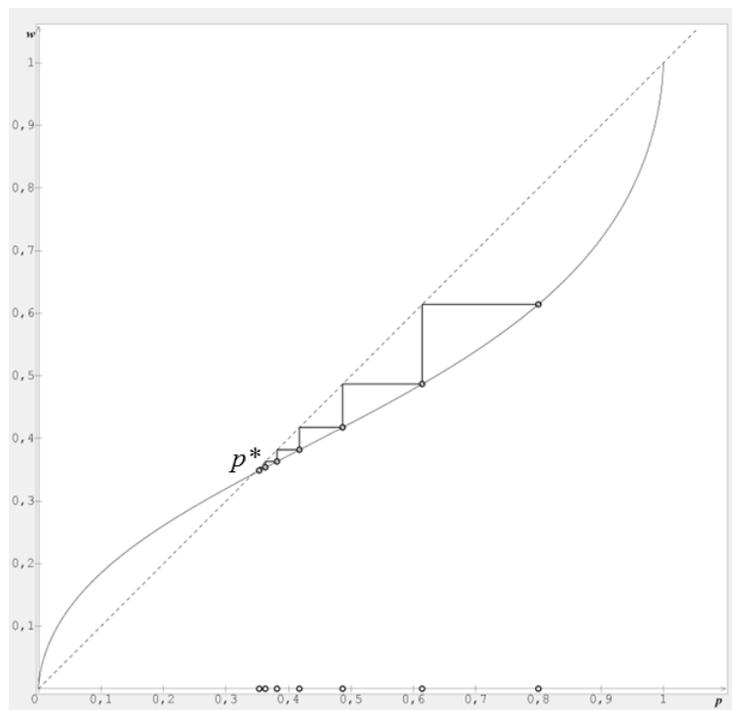

**Fig. 9.** $p_{n+1} = w_\gamma(p_n)$ for $\gamma = \varphi^{-1}$ and $p^* < p_0 < 1$

According to the figure 9, above $p^*$, any series of decisions relying on $p_0$ such that $p^* < p_0 < 1$ shows a decrease in $w_\gamma(p_n)$ down towards $p^*$ what reveals the subjective trap which threatens any agent's hope and motivation to reach a position or a market share associated with a high probability [of success].

Considering the global experiment led by Tversky and Kahneman, an easy interpretation could be that members of WEIRD societies are realistic only for certainties – $p_0 = 0$ or $p_0 = 1$ – or if they are left to their own reasoning while they are involved in an activity which requires decision-making, their sense of reality coincides with $p^*$: although reality cannot be reduced to a sole notion – as welfare relies not only on an income – it is understandable that what is real for a human being is what he or she is motivated for according to various levels of needs which depend on any agent. Involvement and motivation imply the will to structure a field of activity or a matter which constitute the reality of the agent among all the other agents of a society who also develop their own, but sometimes shared, "real world". So in an environment which is in a perpetual evolution, motivation – involvement and will – is, through decision-making and action, the reality of tomorrow and for WEIRD people it can be proposed that it is quantifiable at the average level of $p^* \approx 0.34$ at the end of a business cycle.

Furthermore, the notion of reality is also a compromise between the expectations of an agent involved in an exchange and the good or service or money he or she gets from this exchange. But the satisfaction drawn from it depends on the initial level of certainty to get the expected product and above all on the relative positions of $p_0$ and $p^*$.

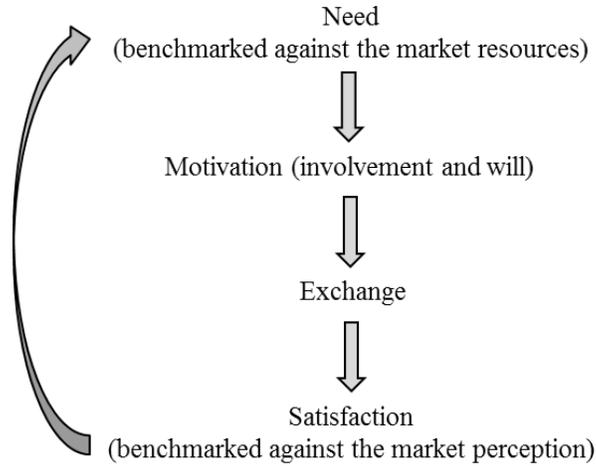

**Fig. 10.** Qualitative Business Cycle (QBC)

If $p_0 > p^*$, the real probability for the transaction to take place is below the one estimated by the agent: the prospect of gain through the exchange is followed by the feeling that this gain has a lower perceived value than anticipated. A feeling which is predominant when time goes by until the end of the business cycle (figure 10), when $p^*$ is reached, along which the agent goes from need expression to involvement in the exchange and then to a relative satisfaction that is even less important since the market is known from inside as an easy outlet from a probabilistic viewpoint. The bargain proves to be less interesting than expected although it can be surprising by its relative ease. While the outcome of the transaction stays the same during the business cycle, the gap $p_0 - p^*$ characterizes the subjective loss of value or the level of overinvestment. As a consequence, a WEIRD agent who is confronted with this kind of situation, knowing this effect by experience, may be hesitant to commit himself in such a transaction: he then plays down the highest real probabilities, except 1, and adopts an attitude that can be interpreted as laziness.

If $p_0 < p^*$, the real probability for the transaction to take place is above the one estimated by the agent: the prospect of gain through the exchange is followed by the feeling that this gain has a greater perceived value than anticipated. This feeling goes with the business cycle until its end when reality emerges as being a profitable exchange, the estimated probability for it to occur being sufficient while the probability of reference is higher. After taking part in the market, the agent discovers that through his or her involvement, he or she made a profit – which is the product of the outcome by $(p^* - p_0)$ – on the exchange with a low probabilistic "investment".

The disappointment or the satisfaction has for an origin the natural evolution of markets and the feeling of a relatively successful bargain according to the utility of goods or services – or to the level of gain –, objects of transactions, after a while within successive business cycles and with the agent's understanding of his or her relative position on markets.

Eventually, welfare – or perception of welfare – relies on the mechanism described above and settles during successive exchanges thought as being more or less profitable according to probabilities for them to take place. So $p^*$ is an indicator of well-being: it is the point to which those probabilities converge in WEIRD societies and at which neutral motivation and commitment in a market enable to live a kind of hedonism.

*6.2. Poor Economics*

In order to extend the Prospect Theory, which applies originally to WEIRD societies, to the subject of research by Duflo and Banerjee (2012), the weighting operator can be used also in considering the case of negative maturities and corresponding real canonical solutions that may be associated with the economic context of developing countries or with "poor economics".

So for $M_s, M_d < 0$ and $\gamma = \varphi\kappa^{-1}$ the anticipated probability weighting bias is cancelled for $\gamma = 1$ that is to say for:

$$\frac{|M_s|}{c_s} = \frac{|M_d|}{c_d} = \kappa^{-1} = \varphi^{-1} \tag{100}$$

The absolute value of the maturities enables to conceive $\varphi^{-1}$ as a probability that could be reached by the weighting operator within the "poor economics" framework but what follows leads to consider that any reality associated with $\varphi^{-1}$ is probably the one of hope more than of an attainable aim.

When $\kappa = 1$ and $\gamma = \varphi$, the weighting operator is shown in the following figures where $p'*$ satisfies $p'* = w_\gamma(p'*), 0 < p'* < 1$.

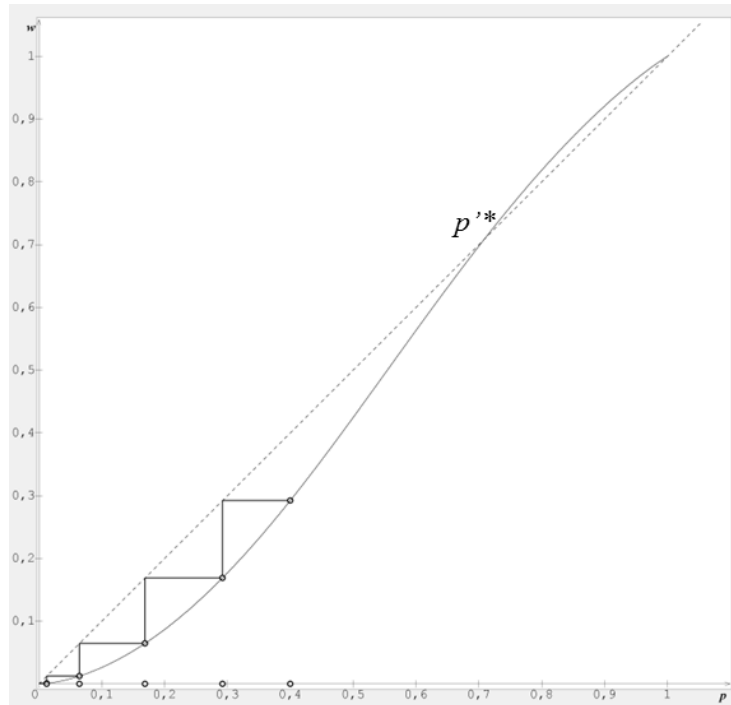

**Fig. 11.** $p_{n+1} = w_\gamma(p_n)$ for $\gamma = \varphi$ and $0 < p_0 < p'*$

The figure 11 illustrates the poverty trap which poorest people in the world are faced with. Below $p'^*$, any series of decisions relying on $p_0$ such that $0 < p_0 < p'^*$ shows a decrease in $w_\gamma(p_n)$ down towards 0 what reveals the more or less subjective trap which threatens any agent's hope and motivation to reach a situation associated with a probability below $p'^*$. This poverty trap means that whatever the initial motivation characterized by $p_0$ below $p'^*$, any involvement of the agent leads to a moral state or position where nothing is possible: the Qualitative Business Cycle (QBC) may still apply but at its end, whatever the satisfaction, no benchmark against any market perception is conceivable probably because of the impossibility of elaborating a market awareness whether it is due to [a lack of] education and society running tradition or to adherence to a particular way of life like a "survival mode". What can be interpreted as laziness in WEIRD societies when $p_0 > p^*$ may then be explained as being inevitability, the main feeling of the poorest when they are trapped into poverty $(p_0 < p'^*)$.

An intermediate conclusion is that the simple concept of market is the first step towards sharing amongst the poorest people a means of building a strategy of exchange of goods or services and thus of growing although in some circumstances such an idea may not survive a long time against inescapable constraints or disasters: one strategy is enough to enter a market (Chauvet 2013) while creating it by the sole strength of a thought and an intention or motivation.

In such a situation any hope is an opportunity for the agent to try to reach $p'^*$, the level of reality that stays as unstable as a dream which may end in starvation and lack of medical care.

So the three "anchorage" points of reality for the poorest are $p = 0$, $p'^* \approx 0.71$ and, beyond any dream, $p = 1$. The first one is reached for more than 70% of the probabilities conceivable as the first will of a "poorest's day". The second one may be an aim but is unstable and cannot be attained precisely except as a result of a lottery draw.

The third "anchorage" point, shown in figure 12, is unavoidable for most people who belong to a country: societies are formed of groups of individuals who obey a hierarchy the structure of which imposes the attributions they "deserve" because of their birth or of their personal proven value – their work, experience, etc. – or of their election by their fellow citizens. Whatever the structure, it generally gives a great place to oligarchies [or castes] legitimated by birth, ideology or religion, education, compliance with some rules or laws, wealth, etc.

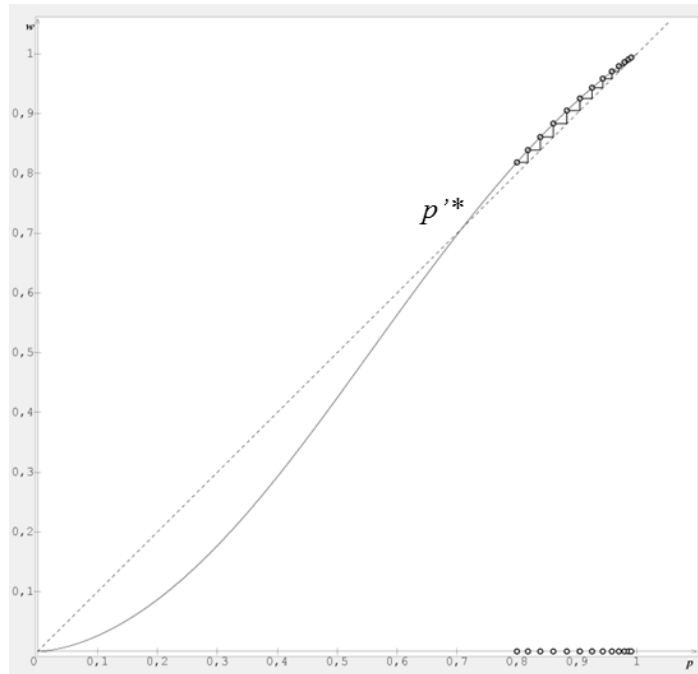

**Fig. 12.** $p_{n+1} = w_\gamma(p_n)$ for $\gamma = \varphi$ and $p'^* < p_0 < 1$

Consequently, less than 30% of the [upper] probabilities conceivable as the initial level of involvement of citizens from the poorest countries lead to a reality of fulfilment according to a business cycle the end of which is satisfaction on a market perceived as such by the privileged.

So, in the societies where the canonical real solutions with negative maturities apply, the weighting operator shows inequality and a strong bipolarity of what can be considered as realities.

*6.3. The Systemic Interpretation*

Daniel Kahneman (2011) went into the dichotomy between two ways of thinking, System 1 and System 2: "System 1 operates automatically and quickly, with little or no effort and no sense of voluntary control," whereas "System 2 allocates attention to the effortful mental activities that demand it, including complex computations." According to this psychologist, the weighting is an operation of System 1. This operation is shown on the figure 13 as the arrow from the abscissa to the curve of the weighting operator and then as the vertical arrows.

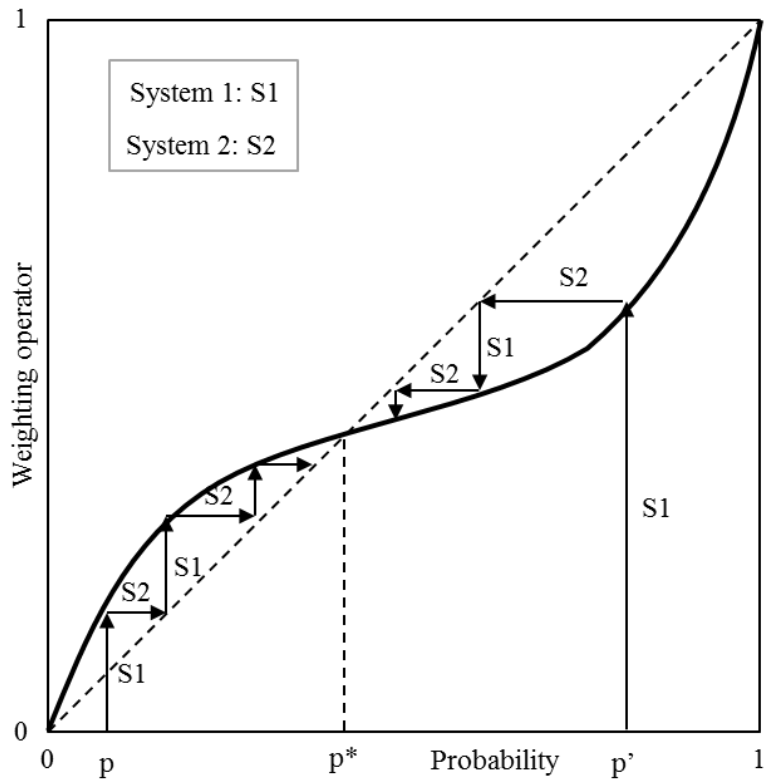

**Fig. 13.** WEIRD System 1 and System 2 processes

As Kahneman wrote, "System 2 articulates judgments and makes choices, but it often endorses or rationalizes ideas and feelings that were generated by System 1." So S1 leads to a first weighting which is the result of past experiences, and memories of comparable situations, but also of biases that affect the agent's cognition. Then S2 tends to compute a rational justification of this first assessment so as to reach an unbiased position relying on the weighting value. This process of coupling of System 1 and System 2 is repeated until both systems "find an agreement" at a convergence point which characterizes the mental equilibrium of the agent with the corresponding temporary relative satisfaction.

Naturally, the poor weighting can also lead to other equilibria through contextual justifications, the existence of two Systems in the mind being universal and finding probably psychophysiological origins in the human cerebral hemispheres specificities.

The philosophy of this model is then that whatever the psychological trap, the 'laziness' one in WEIRD societies, or the poverty one, it relies on an internal process that can be broken down thanks to the intervention of external agents through education and teachings or management best practices based on benevolence and aiming at a mutual growth.

## 7. Growth sustainability

The sets of functions $F$ and $F_R$ are discreet sets and so are the sums of the corresponding adaptive utilities or works or energies that constitute the main physical features chosen to describe the associated functionings and their levels of positive or negative externalities $E_i$'s and $E_i^R$'s. Both couples $(F_s, F_R)$ and $(F_d, F_R)$ are used in order to calculate the maturities $M_s$ and $M_d$ which are then, at first sight, dependent on $\lceil m_s \rceil$ and $\lceil |m_d| \rceil$ respectively. Nevertheless those dependences are not strict and may be factually unverifiable while taking account of the relations between $m_R$, $\lceil m_s \rceil$ and $\lceil |m_d| \rceil$. Considering "ideal" engineering and industrial processes, the maturities should be unitary, their difference from 1 should then be marginal and $m_R$ should be such that $m_R = \lceil m_s \rceil = \lceil |m_d| \rceil$. In a less ideal case, it seems clear that project hazards can involve a kind of more or less controlled maturity randomness which neither depends on $m_R$, $\lceil m_s \rceil$ nor $\lceil |m_d| \rceil$. So, integrating the equation (69), or its equivalent for the demand, can be carried on considering the maturity as a constant. The global capital can be calculated and the capital growth is then written, when the exchanged maturity and complexity are homogeneous between the supply and the demand:

$$\Delta K = K - K_{O(s,d)} = \frac{c}{M} \rho^* \left[ \ln\left(\frac{\rho_{Od}}{\rho_{Os}}\right) + M_{Os} \frac{m_{Os}}{c_{Os}} + M_{Od} \frac{m_{Od}}{c_{Od}} \right] \qquad (101)$$

Where:

$$\frac{c}{M} = \frac{c_s}{M_s} = \frac{c_d}{M_d} \qquad (102)$$

The condition for capital growth to be effective within a transaction is still $\Delta K > 0$ but the introduction of maturities shows a new complexity for the analysis of this global growth condition.

Both factors of $\Delta K$, the current exchange part relying on the triplet $(c, \rho^*, M)$ and the legacy part taking account of previous transactions, must be studied.

$$r_M = \frac{\rho_{Od}}{\rho_{Os}} > \exp\left(-M_{Os} \frac{m_{Os}}{c_{Os}} - M_{Od} \frac{m_{Od}}{c_{Od}}\right) \qquad (103)$$

The generic maturity definition is:

$$M = \frac{\sum_{i=1}^{m_R} E_i^R}{\sum_{i=1}^{m} E_i} \quad (104)$$

which can be instantiated for each specific maturity $M_s$, $M_d$, $M_{Os}$, and $M_{Od}$ with the notations introduced previously in this article. The marketing part of $M$ is the denominator which is always strictly positive: commercially, it is necessary for the supplier to present a positive view of the product he or she wants to provide and sell, while the purchaser needs to believe that what she or he is ready to buy has a global positive [adaptive] utility. As for the numerator of $M$, it must be positive at the time of the purchase for a sustainable reputation of the supplier but may turn into a global negative externality after a while because of the obsolescence of the exchanged system components: this obsolescence can be intrinsic, due to the system aging and attrition, or it can be declared so because of the evolution of the market standards and norms.

### 7.1. The virtuous circle

When all maturities are positive, the initial reasoning about the conditions of growth and inflation applies simply in replacing the original c's by the ratios $c/M$ for the reference frames and for the current exchange. The growth is then all the more effective since $r_M$ is high with the same conclusions, on the relative levels of time cots for the demand and the supply, as originally proposed.

But introducing maturity in the current exchange shows that a higher maturity reduces the increase in capital: growth is kept at a sustainable level in the considered exchange which provides a new reference frame for future transactions wherein the high maturity will induce a higher margin of growth for the next exchange when $r_M$ is fixed and according to equation (101). All other parameters equal, high maturities involve a virtuous circle whereby growth is built on the capitalization of previous gains. Nevertheless, it is important to notice that great complexities can alter this positive effect: high a c in the current exchange enables a higher growth but can lead to future recessions unless complexity and demand [time cost] increase steadily.

*7.2. The erroneous growth*

When all maturities are negative and $r_M$ is kept below the exponential term's value, growth exists but only from a superficial point of view: the increase in capital is obtained at the expense of the exchanged system environment because of negative externalities. The problem with such a growth is that its consequence, the alteration of the ecosystem, that it be natural or man-made, cannot be stated until the harm does not occur. And by the game of power or influence of lobbies or other sovereign authorities, the harmful activities can be as sustainable as a virtuous circle.

*7.3. Is a transition to virtuous growth possible?*

Beyond this bipolarity of growth based on clear negative or positive values of maturities and associated externalities, a transitional path can be found from erroneous to virtuous growth. It consists in increasing the demand [time cost], expecting its better civic behavior after learning positive, ethical values and accountability; and lowering the supply of harmful products, through markets regulations by international and independent authorities and transferring the labor forces from those products production lines to sustainable virtuous growth firms. But such mechanisms may lead to both high inflation and unemployment rates with no certainty of [virtuous] growth during the transition. This prospect is sufficiently compelling for the sovereign authorities of countries with erroneous growth to refuse to engage in such attempts to save the diversity of species and the natural "assets" of planet Earth or to go against lobbies and other pressure groups or sources of economic power on the world markets.

## 8. Conclusions

### 8.1. Conclusions drawn from canonical solutions

The lesson drawn from the paragraph "Industrial processes under estimates" is that any business is worth being developed: it creates or maintains employment, ensures activity for providers and solutions to their customers who run their own businesses, what is a well-known virtuous circle, the basis of which are exchanges according to the law of supply and demand, formalized or not by contracts. That seems evident in WEIRD societies where there is so high a density of such exchanges that the markets to which they give a concrete expression are an inescapable and well-structured reality, from Stock Exchange and trade to simple shopping. Whereas in the poorest societies, material resources may be too scarce to generate a persistent market awareness meeting the WEIRD standards in spite of other kinds of trade customs, often founding a barter economy.

Nevertheless, this statement of fact relies on a model which is WEIRD although it represents an attempt at including into the same economic system a great deal of behaviors of every kind of society. The canonical real solutions of the market dynamics equations enable the description of patterns which apply to WEIRD societies and to the poorest ones according to maturities of markets. But the canonical notion leads naturally to enlightening every situation that cannot be reached because of too high a complexity level. Thus, a first step towards complexification is to understand that both canonical real solutions for positive and negative maturities coexist in a great deal of countries where all kinds of societies can be found and studied, even if some of them are a matter for the WEIRD characterization and others meet the poorest reference.

### 8.2. Out of the Canonical Solutions

The market dynamics model is built on clear principles of value and integrates elementary notions of labor time cost, physical work or adaptive utility, capabilities and agent – following Sen (2001) –, work utility, capacity and payoff. The concepts associated with these notions rely on a formalism the equations of which are completely coherent with one another. But the great number of parameters and variables that create complexity and paradoxically enable its control, require arbitrary decisions guided by the choice of a mythical mathematical reference: The Gold Number. Out of this choice and out of the equilibrium between payoff and value, the market dynamics is not really chaotic because other decisions can be made that lead to more complex "non-canonical" solutions but the perfect fit between supply and demand remains exceptional for finding connections with behavioral economics. Further use of the market dynamics model based on computer programs could also bring to light non-trivial conclusions that may not be reached either by fast, automatic operations of System 1 or slower, controlled operations of System 2 mind "processes."

### 8.3. Exchange, Learning and WEIRD Market

Any exchange according to the law of supply and demand creates a global capital growth provided that it goes with a pedagogical attitude from the supplier towards the buyer who must accept to spend time on getting used to the more or less complex purchase he or she makes. An exchange comes within a necessary "learning" duration which counts much in the processes of growth. This makes education the main foundation of the economic successes. Nevertheless, schools or universities should not be and are often not the only mainstays of the upper standards of societies. They are one of the steps that lead to a life of responsible man or woman whose behaviors show a strong involvement in ethical decision-making and action. Ethics is the cornerstone of the consistency of a society and may find a kind of universal way of behavior applicable in any economic context while understanding the mind-body duality that philosophically governs human decisions (Chauvet 2016). But learning ethics does not guarantee such universality, physical constitution and mental processes forging unicity and rivalries or sometimes developing disabilities (Chauvet 2018) that prevent from being receptive to some kinds of education, teachings or trainings and then lead to social exclusion. Inequalities amongst people may be basically ethical according to the way they are dealt with in societies where values are standardized and behaviors, criteria for the selection of the most appropriate candidate for a position. The law of supply and demand is a means for everyone of finding equilibrium between needs, capabilities and the requirements of markets. But it is still necessary to have access to a market and at least to understand the conditions of a transaction or of a profitable exchange of good or service, realities that may be not permanent or sustainable among the poorest people.

The role of superior institutions and the person-to-person transmission of tacit knowledge through apprenticeship historically explain a part of the processes whereby Western Europe has entered the industrial age while other world regions were showing less demographic, technological and financial successes (de la Croix et al. 2018). History enables such theses, especially brilliant when they are supported by sound models, which confirm the importance of the skill transmission in the economic growth. But those theories are anchored in the evolution and can be strengthened while considering that one of their topics is the concern of the most common economic law. A functional approach to basic economics enables a fine structuring of processes in use on markets: labor forces can be modelled and integrated within a capitalistic description of the law of supply and demand for eventually enlightening the deep connections between growth and "knowledge how" transfer associated with the trade of goods and services or more generally of complex systems.

On the one hand, in the light of this study, educational institutions, from the family considered as such, to schools and universities, appear as the first places where ethical behavior and performance can be rewarded but where education or teachings are only intended to enable the creation or the management of a family with love and benevolence, or to be used to learning how to learn or to teach efficiently. So, these institutions enable to enter the markets of [future] husbands and fathers, wives and mothers, or of teachers. Nevertheless, universities also trade on the knowledge market and through research and a hard competition for the access to the highest academic distinctions, they provide a notable part of the growth of a society in compliance with the model of this article.

On the other hand, the vast majority of the citizens of a country cannot enter this market because, like the poorest, they do not even conceive or imagine its mechanisms which are not only based on an initial investment, or a kind of purchase, but also on a long lasting effort spent on exchanging know-how and inter-personal skills. So those left behind by educational institutions get into the labor market with a relative success relying on their capabilities of learning by doing and finding in their colleagues' or managers' behaviors a relatively sustainable understanding of personal growth dynamics. Their involvement in projects is sometimes the opportunity to be trained on new technologies or practices but the less their time cost the less they are likely to contribute to a global growth; thus the less they are valued and at the end of a business cycle they find themselves [back] at the WEIRD reality point while they are maintained in a feeling of a low profit, if not a deficit, after underestimating the probabilities of a world of opportunities. When their time cost becomes high enough to generate growth within a skill acquisition process, while their capabilities are important and so that they progress in a great uncertainty, they tend to overestimate probabilities of prospects and to show a high motivation: they represent a valuable investment for their employer and they reach the WEIRD reality point at which they may have the impression of making a real profit.

### *8.4. The Paradox of the Poorest Capitalism and WEIRD Labor Forces*

In harsher countries and societies which go under tragic circumstances, failures, corruption or bankruptcies, this point may be vanishing and replaced by the poorest "anchorages". But such harshness can also be the result of a fierce competition which, although generally respecting ethics and laws, is led by these leaders who make of their reality the one of thousands or even millions or billions of people after manufacturing devices and developing digital media that support their visionary conception of markets. These magnified agents' views have a singular property: they shape the economic world and they quickly become inescapable within the necessity of a successful business. Furthermore, they crystallize the collective imagination of the consumers on sets of functions combined into a single product and in its following range while offering solutions to the need that they create. Those specific agents show the gap between them and less successful businessmen when the poorest weighting model is applied to a capitalistic society. While a few visionary leaders escape from the poverty trap and drive their firms towards a reality which is associated with the certainty of a great market share, other entrepreneurs fall down to the disappearance of their business.

Nevertheless beyond the always temporary certainty on a market, any business is subject to unknown factors: seminal ideas may be shared or stolen and successfully exploited by other visionaries than their originators; patents can be circumvented by a contender for the production of more competitive goods; the business or economic model of a service provider can be copied and enhanced by a competitor and so on. Thus the marketing strategies of a firm must be found as the equilibrium between official information which is made free for catching the costumers and confidentiality of the technical functions, knowing that once the product is available on the market, it becomes the property of any buyer who can analyze and reproduce it while elaborating and using other patents than those already in force. So, in a harsh competition what matters is time: time to market, time to deliver, lifetime of an asset and eventually time to obsolescence. When leaders work on those strategies, they may have certainties to reach the highest probability of market efficiency. If they had not, they would not be able to manage their teams convincingly and their transfer of uncertainties to the designers, engineers and project, operation or production managers, would not be understood as a source of motivation, involvement and will beyond any contractual commitment. So, within contracts and tacit agreement of responsibility sharing for the success of the firm, a project – that may be led in coordination with other industrial companies, possibly subcontractors – is managed under time constraints and taking account of organizational and technical complexities in order to satisfy the customer according to a schedule and a budget. The matrix formalism allows to trace the initial estimate "down" to the technical achievement of the contractual functions and underlines the mechanism whereby externalities may appear, especially those which are the concern of obsolescence management. The trace followed reveals also that uncertainty is transferred from the top of a hierarchy to the technical teams that have to deal with possible externalities or obsolescence according to a policy elaborated in agreement with the upper management directives. These teams, or labor forces, may be essentially WEIRD, even in the poorest but industrialized countries, while their leaders are the select few on the top of the poorest model applied to the capitalistic society where there can be also poor people in despair. But those privileged, except if they break the law, are highly protected by their wealth which enables them to gamble with a high probability of gain, thanks to their imagination, their innovation spirit, their knowledge of their target markets and their understanding of how to motivate labor forces.

# APPENDIX A

Let $P = \left[ p_{ij} \right]_{1 \leq i, j \leq n}$ be a permutation matrix of dimension n, for instance n=3:

$$P = \begin{pmatrix} 1 & 0 & 0 \\ 0 & 0 & 1 \\ 0 & 1 & 0 \end{pmatrix}$$

$P$ satisfies the following conditions (permutation matrix characterization):

$$\sum_i p_{ij} = 1; \quad \sum_j p_{ij} = 1; \quad \sum_i \sum_j p_{ij} \ln p_{ij} = 0$$

Let $B$ be the matrix built from $P$ in replacing each 1 in $P$ by a bistochastic matrix $B_{ij}$:

$$B = \begin{pmatrix} B_{11} & (0) & (0) \\ (0) & (0) & B_{23} \\ (0) & B_{32} & (0) \end{pmatrix}$$

$B$ is also bistochastic. According to the theorem of Birkhoff – von Neumann $B$ can be decomposed in permutation matrices $P_\alpha$ such that:

$$B = \sum_\alpha C_\alpha P_\alpha, \quad \sum_\alpha C_\alpha = 1$$

Where $P_\alpha$ has necessary the form:

$$P_\alpha = \begin{pmatrix} P_{11}^\alpha & (0) & (0) \\ (0) & (0) & P_{23}^\alpha \\ (0) & P_{32}^\alpha & (0) \end{pmatrix}$$

Proof: clearly by contradiction.

Furthermore $P_{11}^\alpha, P_{32}^\alpha, P_{23}^\alpha$ are also permutation matrices and:

$$B_{ij} = \sum_\alpha C_\alpha P_{ij}^\alpha$$

# APPENDIX B

The equation:

$$\kappa_{Os} \frac{d}{dm_s}\left[\frac{d\rho}{dm_s}\sum_{i=1}^{i=\lceil m_s \rceil} E_i^s\right] = \frac{d\rho}{dm_s}\sum_{i=1}^{i=\lceil m_s \rceil} E_i^s - \frac{1}{c_s}\sum_{i=1}^{i=m_R} E_i^R \rho$$

Becomes, with $m_{\lceil s \rceil} = \lceil m_s \rceil$:

$$\kappa_{Os} \frac{d^2\rho}{dm_s^2}\left[\sum_{i=1}^{i=m_{\lceil s \rceil}} E_i^s\right] + \kappa_{Os}\frac{d\rho}{dm_s}\left[\frac{d}{dm_s}\sum_{i=1}^{i=m_{\lceil s \rceil}} E_i^s\right] = \frac{d\rho}{dm_s}\sum_{i=1}^{i=m_{\lceil s \rceil}} E_i^s - \frac{1}{c_s}\sum_{i=1}^{i=m_R} E_i^R \rho$$

Where

$$\frac{d}{dm_s}\sum_{i=1}^{i=m_{\lceil s \rceil}} E_i^s = \frac{dt}{dm_s}\frac{d}{dt}\sum_{i=1}^{i=m_{\lceil s \rceil}} E_i^s(t)$$

Whether it is because of the law of conservation of the energy or due to a budgetary constraint (t is time):

$$\frac{d}{dt}\sum_{i=1}^{i=m_{\lceil s \rceil}} E_i^s(t) = 0$$

While innovation is a necessity for the supplier to keep up with the perpetual economic evolution:

$$\frac{dm_s}{dt} > 0$$

And finally

$$\kappa_{Os} \frac{d^2\rho}{dm_s^2} \left[ \sum_{i=1}^{i=m_{\lceil s \rceil}} E_i^s \right] = \frac{d\rho}{dm_s} \sum_{i=1}^{i=m_{\lceil s \rceil}} E_i^s - \frac{1}{c_s} \sum_{i=1}^{i=m_R} E_i^R \rho$$

The same reasoning applies to work out the fundamental equation of market dynamics for the demand.